\documentclass[twocolumn,a4paper,reprint,amssymb,aps,prd,groupedaddress,superscriptaddress,amsmath,nobibnotes,showpacs,nofootinbib]{revtex4-2}

\usepackage{graphicx,bm,color,psfrag}
\usepackage{amsfonts}
\usepackage{lipsum}
\usepackage{enumitem}
\usepackage[dvipsnames]{xcolor}
\usepackage[colorlinks = true,
            linkcolor = blue,
            urlcolor  = blue,
            citecolor = blue,
            anchorcolor = blue]{hyperref}

\usepackage{soul}
\usepackage{xcolor}
\usepackage{threeparttable,booktabs}

\begin{document}
\title{Quadratic energy-momentum squared gravity: \\constraints from big bang nucleosynthesis }

\author{\"{O}zg\"{u}r Akarsu}
\email{akarsuo@itu.edu.tr}
\affiliation{Department of Physics, Istanbul Technical University, Maslak 34469 Istanbul, Turkey}

\author{Mariam Bouhmadi-L\'opez}
\email{mariam.bouhmadi@ehu.eus}
\affiliation{IKERBASQUE, Basque Foundation for Science, 48011, Bilbao, Spain.}
\affiliation{Department of Physics and EHU Quantum Center, University of the Basque Country  UPV/EHU, P.O. Box 644, 48080 Bilbao, Spain.}

\author{Nihan Kat{\i}rc{\i}}
\email{nkatirci@dogus.edu.tr}
\affiliation{Department of Electrical-Electronics Engineering Do\u gu\c s University, \" Umraniye 34775 Istanbul, Turkey}

\author{N. Merve Uzun}
\email{uzunmer@itu.edu.tr}
\affiliation{Department of Physics, Istanbul Technical University, Maslak 34469 Istanbul, Turkey}

\begin{abstract}
In this work, we extend the standard cosmological model within the quadratic energy-momentum squared gravity (qEMSG) framework, introducing a nonminimal interaction between the usual material field ($T_{\mu\nu}$) and its accompanying partner field (qEMSF, $T_{\mu\nu}^{\rm qEMSF}$), defined by $f(\mathbf{T}^2) = \alpha \mathbf{T}^2$ with $\mathbf{T^2}=T_{\mu\nu}T^{\mu\nu}$. Adopting an analytical approach within the qEMSG framework, we present a comprehensive exploration of Big Bang Nucleosynthesis (BBN) dynamics. Our analysis selects the radiation-dominated universe solution compatible with the standard cosmological model limit as $\alpha \rightarrow 0$ and reveals that qEMSF interaction model can modify the radiation energy density's evolution, potentially altering neutron--proton interconversion rates and consequently affecting $^4$He abundance in various ways. By explicitly defining modifications to the predicted primordial $^4$He mass fraction, $Y_{\rm p}$, we establish the most stringent cosmological constraints on the parameter $\alpha$ based on recent measurements of $Y_{\rm p}$: $(-8.81 \leq \alpha \leq 8.14) \times 10^{-27} \, \mathrm{eV}^{-4}$ (68\% CL) from Aver \textit{et al.}'s primordial $^4$He abundance measurements, aligning with $\alpha=0$. Additionally, $(3.48 \leq\alpha \leq 4.43)\,\times 10^{-27} \rm{eV}^{-4}$ (68\% CL) from Fields \textit{et al.}'s estimates, utilizing the Planck-CMB estimated baryon density within the standard cosmological model framework, diverges from $\alpha=0$, thereby lending support to the qEMSF interaction model. The study also highlights the bidirectional nature of energy-momentum/entropy transfer in qEMSF interaction model, depending on the sign of $\alpha$. The implications of qEMSF in the presence of additional relativistic relics are also explored, showcasing the model's potential to accommodate deviations from standard cosmology and the Standard Model of particle physics.
\end{abstract}

\maketitle

\section{Introduction}
\label{sec:intro}
Generalizing the matter Lagrangian density in a nonlinear way, for instance, by including an analytic function of the Lorentz scalar ${\bf T^{2}}=T_{\mu \nu }T^{\mu \nu }$~\cite{Katirci:2014sti,Roshan:2016mbt,Akarsu:2017ohj,Board:2017ign} derived from the energy-momentum tensor (EMT), $T_{\mu \nu }$, introduces new contributions of material stresses to the right-hand side of the Einstein field equations. It has recently been shown in Ref.~\cite{Akarsu:2023lre} that such generalizations, which modify the introduction of the material source in the conventional Einstein-Hilbert  action by adding arbitrary functions of matter-related scalars (like $f(\mathcal{L}_{\rm m})$~\cite{Harko:2010mv}, $f(\mathcal{T})$~\cite{Harko:2011kv} ($\mathcal{T}$ is the trace of the EMT), and $f(\bf T^{2})$~\cite{Katirci:2014sti,Roshan:2016mbt,Akarsu:2017ohj,Board:2017ign}) to the usual matter Lagrangian density $\mathcal{L}_{\rm m}$, are equivalent to general relativity (GR) with nonminimal matter interactions. The fact that the gravitational part of the Lagrangian density remains unchanged renders these models phenomenologically interesting because they differ from the usual dynamical dark energy/modified gravity models in that the latter add further dynamical degrees of freedom to the Lagrangian density, often in the form of scalar fields.

Among these matter-type models, energy-momentum squared gravity (EMSG) includes $T_{\mu \nu }$, the usual material field, interacting generally nonminimally with its accompanying partner, the \textit{energy-momentum squared field} (EMSF), which has a unique form constructed via the function $f(\bf T^{2})$. Owing to the nonminimal nature of the interaction between the material field and its EMSF partner, both fields individually violate the local energy-momentum conservation law, leading to $\nabla^{\mu}T_{\mu\nu}=- \mathcal{Q}_{\nu}=-\nabla^{\mu}T_{\mu\nu}^{\rm EMSF}$, where $ \mathcal{Q}_{\nu}$ is the interaction kernel that governs the energy transfer rate between the material field and its EMSF partner. A notable example within this framework is the energy-momentum powered field (EMPF) interaction model, which is formed by selecting the function $f(\bf T^{2})=\alpha({\bf T^{2}})^{\eta}$, where $\alpha$ and $\eta$ are constants~\cite{Akarsu:2017ohj,Board:2017ign}. In a cosmological context, when $\eta<1/2$, the terms arising from the EMPF become significant at low energy densities, thus becoming relevant to the dynamics of the late Universe. For example, by setting $\eta=0$, the background dynamics of the $\Lambda$CDM model can be replicated, and $w$CDM-like extensions of the $\Lambda$CDM model can be achieved for $\eta\simeq 0~$\cite{Akarsu:2017ohj}. Specifically, Ref.~\cite{Akarsu:2017ohj} demonstrates that an $\eta$ value close to zero can explain the late-time acceleration of the Universe due to the contribution of dust from EMPF---since it resembles a cosmological constant. The observational analysis in the work suggests $\eta=-0.003\pm0.023$ at 95\% confidence level (CL). A more recent study~\cite{Kolonia:2022jje} reports $0<\eta<0.18$ at 95\% CL and notes that when the inclusion of a cosmological constant $\Lambda$ alongside EMPF is allowed, $\eta=0.26\pm0.25$ at 95\% CL.

Conversely, for $\eta>1/2$, the EMPF becomes more prominent at high energy densities, suggesting that its effects could be observed in the early Universe and within dense compact astrophysical objects, like the high-density cores of neutron stars. The case $\eta=1$ of the EMPF is the most straightforward form of the EMSF interaction model, known as the quadratic energy-momentum squared field (qEMSF), namely, $f({\bf T^{2}})= \alpha {\bf T^{2}}$~\cite{Roshan:2016mbt,Akarsu:2017ohj,Board:2017ign}. This form yields Friedmann equations incorporating second-order matter terms that are reminiscent of the
terms (corrections) arising in the type II Randall–Sundrum brane cosmology 
\cite{Randall:1999vf,Csaki:1999jh,Cline:1999ts,Binetruy:1999hy}, more generally in brane-world cosmological constructions~\cite{Brax:2003fv}, for $\alpha>0$ and in loop quantum cosmology~\cite{Ashtekar:2011ni} for $\alpha<0$, though the underlying physics of these models significantly differ.  Barrow~\cite{Barrow:2020tzx} has proposed that quantum-gravitational effects might introduce complex, fractal features on black hole horizons, leading to cosmological theories with field equations akin to those in EMSF interaction model.  Particularly, Barrow entropy correction to the Friedmann equation assumes a quadratic form of energy density for the most intricate and fractal horizon structure ($\delta=1$)~\cite{Sheykhi:2021fwh}. It has been demonstrated in~\cite{Chen:2019dip,Chen:2021cts} that $f({\bf T^{2}})$ interaction\footnote{While considering the Maxwell field within the EMSG (or EMSF) framework, in Refs.~\cite{Roshan:2016mbt,Chen:2019dip,Chen:2021cts}, authors include the second metric derivative of the matter Lagrangian density in their analysis. However, according to the common recipe of EMSF, this term should be excluded also for material fields other than perfect fluid, even if it does not vanish, see Section~\ref{sec:EMSG} and also Ref.~\cite{Akarsu:2023lre} for details.} might violate the geometric correspondence between high-frequency (eikonal) quasi-normal modes and photon geodesics (ring) around black holes, potentially serving as a distinct indicator for models extending beyond GR with minimal interaction. In Ref.~\cite{Akarsu:2018zxl}, the authors constrained $\alpha$, which quantifies the amount of qEMSF relative to the usual material field, to $|\alpha|\lesssim 10^{-35} \,{\rm eV^{-4}}$. This constraint was derived by comparing computed mass-radius relations for neutron stars, using realistic equation of state (EoS) parameters, with observed relations for actual neutron stars. For a similar astrophysical constraint, refer also to Ref.~\cite{Nazari:2022xhv}, which reports $|\alpha| \lesssim 10^{-35} \, \mathrm{eV}^{-4}$ based on binary pulsar observations. When the tightest astrophysical constraint is applied in a cosmological context, it ensures that standard cosmology remains largely unaffected up to energy scales even larger than those relevant to the quark-hadron phase transition, which occurs at an energy density of $10^{31}\, {\rm eV^{4}}$ when the Universe is approximately $t\sim 10^{-4}\, \rm s$ old.
In their analysis, $\alpha$ is allowed to have both positive and negative values. Notably, a negative $\alpha$ can lead to an effective EoS stiffer than the Zeldovich fluid $(P/\rho=1)$, which may indicate potential stability issues (see footnote 5 in Ref.~\cite{Akarsu:2018zxl} and for a similar study, see Ref.~\cite{Nari:2018aqs}). Chaotic inflation has been extended via the EMSG, i.e., EMPF interaction model, presenting that when the canonical scalar field is considered as a usual source, it is accompanied by a specific form of K-essence fields~\cite{HosseiniMansoori:2023zop} and this scenario enables the formations of primordial black holes (PBHs) and second-order gravitational waves (GW)~\cite{HosseiniMansoori:2023mqh}. Their analysis has constrained $\alpha$ to negative values to avoid issues of ghost and gradient instabilities.

The era of Big Bang Nucleosynthesis (BBN) serves as an invaluable laboratory, encapsulating physics at energies below $10\, \rm MeV$. This is distinct from the very early stages of the Universe at ultra-high energies and the later stages on ultra-large (length) scales (see Refs.~\cite{bbf,kolb,Olive:1999ij,Cyburt:2015mya,muk} and Ref.~\cite{Baumannbook} for detailed discussions). The synthesis of light elements during BBN is highly sensitive to the conditions prevailing in the early radiation-dominated epoch at around $t\sim1\, \rm s$ (temperatures $T\sim 1\, \rm MeV$), a period marked by the decoupling of neutrinos. Therefore, BBN imposes stringent constraints on deviations from standard cosmology and potential new physics beyond the Standard Model of elementary particles. In the standard cosmological scenario -- assuming the dynamics of the Universe is well described by the standard $\Lambda$CDM model and the Standard Model of particle physics -- the physical processes relevant to BBN commence when the Universe's temperature drops to $\sim 1 \, {\rm MeV}$ and continue until $\sim 0.1\, {\rm MeV}$. During this period, the Universe is populated only by free protons and neutrons in baryonic ingredients, and its age (scale factor) spans from $t\sim 1\,{\rm s}$ ($a\sim 2\times 10^{-10}$) to $t\sim 3\,{\rm min}$ ($a\sim 2\times 10^{-9}$), remaining radiation-dominated. Any viable extension or modification of the standard cosmological model should not significantly alter this scenario. Accordingly, alternative cosmological models constructed within various modified gravity theories, such as $f(R)$ theories~\cite{Faulkner:2006ub}, Brans-Dicke theories~\cite{Akarsu:2019pvi}, teleparallel gravity~\cite{Capozziello:2017bxm}, bimetric theories~\cite{Hogas:2021saw}, $f(Q)$ gravity~\cite{Anagnostopoulos:2022gej}, and nonminimal matter interaction models like $f(R,\mathcal{L}_{\rm m})$ and $f(R,\mathcal{T})$~\cite{Bhattacharjee}, as well as other interesting approaches such as varying gravitational constants~\cite{Alvey}, violating Lorentz invariance~\cite{Lambiase:2005kb}, and Barrow entropy~\cite{Barrow:2020kug}, have been probed and constrained by BBN.  

In this work, we explore constraints on the free parameter $\alpha$, which quantifies the extent of qEMSF relative to the usual material field, stemming from the dynamics of the early Universe, specifically from BBN. Initially, adopting an analytical approach, we conduct a comprehensive mathematical and physical examination of BBN dynamics within the framework of qEMSG. This analysis enables us to explicitly define modifications to the helium-4 ($^4$He) mass fraction, $Y_{\rm p}$. These modifications, in turn, allow us to derive stringent constraints on $\alpha$ using recent astrophysical measurements of primordial $^4$He abundances.\footnote{Another work that imposes constraints on the $\alpha$ parameter of EMSG from BBN perspectives has recently been posted on arXiv by Jang et al.~\cite{Jang:2024jso}. Opting for a numerical approach, this study directly employs the modified Friedmann equation, i.e., $H(z)$, of EMSG along with an updated version of the Wagoner code and revised reaction rates from the JINA REACLIB Database.}

The structure of this paper is organized in the following manner: Section~\ref{sec:EMSG} provides a new interpretation of the energy-momentum squared gravity (EMSG) from the perspective of nonminimal matter interaction. Section~\ref{sec:qEMSF} details the explicit construction of the quadratic energy-momentum squared field (qEMSF) extension of the standard cosmological model. In Section~\ref{sec:thermo}, we apply thermodynamics principles of open systems to the usual radiation component in our model and analyze the direction of energy-momentum transfer, which depends on the sign of the parameter $\alpha$. In Section~\ref{sec:bbn}, we explore the implications of a radiation-dominated universe within the framework of the qEMSF interaction model. We employ a semi-analytical approach to derive modifications in $^4$He abundance, emphasizing its complex connection with changes in the neutron-to-proton equilibrium ratio at freeze-out, as well as with alterations in the timings of both freeze-out and nucleosynthesis. We also constrain the parameter $\alpha$ using the recent estimates of the primordial mass fraction of $^4$He. Additionally, we calculate the necessary constraints on $\alpha$ to align the $^4$He fraction with observed values, considering the presence of one extra relativistic degree of freedom alongside qEMSF. We conclude our findings and discussions in Section~\ref{conclusion}.

\section{EMSG from nonminimal matter interaction perspective}
\label{sec:EMSG}

We begin with the following action constructed by the inclusion of the term $f({\bf T^{2}})$ with ${\bf T^{2}}=T_{\mu \nu }T^{\mu \nu }$ in the usual Einstein-Hilbert action, retaining the bare cosmological constant, $\Lambda$, as per Lovelock's theorem~\cite{Lovelock:1971yv,Lovelock:1972vz}:
\begin{equation}
S=\int {\rm d}^4x \sqrt{-g}\,\left[\frac{1}{2\kappa}\left(R-2\Lambda \right)+\mathcal{L}_{\rm m}+f({\bf T^{2}})\right],
\label{action}
\end{equation}
where $g$ is the determinant of the metric tensor $g_{\mu\nu}$, viz., $g={\rm det} (g_{\mu\nu})$, $R=g^{\mu\nu} R_{\mu\nu}$ is the scalar curvature with $R_{\mu\nu}$ being the Ricci curvature tensor, $\mathcal{L}_{\rm m}$ is the Lagrangian density corresponding to the conventional material field described by the EMT, $T_{\mu\nu}$, and $\kappa=1.69 \times 10^{-37}\, {\rm GeV}^{-2}$ is the Newton’s gravitational constant scaled by $8\pi$, with units such that $\hbar=c=1$.

We vary the action above with respect to the inverse metric $g^{\mu\nu}$ as
\begin{equation}
 \begin{aligned}   \label{variation}
  \delta S=\int\, {\rm d}^4 x \bigg[&\frac{1}{2 \kappa}\frac{\delta R}{\delta g^{\mu\nu}} \sqrt{-g}+\frac{1}{2 \kappa}\frac{\delta (\sqrt{-g})}{\delta g^{\mu\nu}} (R-2 \Lambda)      \\
 & +\frac{\delta(\sqrt{-g}\mathcal{L}_{\rm m})}{\delta g^{\mu\nu}}+ \frac{\delta(\sqrt{-g} f)}{\delta g^{\mu\nu}}\bigg] \delta g^{\mu\nu}.  
\end{aligned}
\end{equation}
Assuming the matter Lagrangian density, $\mathcal{L}_{\rm m}$, depends solely on the metric tensor and not on its derivatives -- an approach applicable to perfect fluids as well as Maxwell, scalar, and gauge fields -- the conventional definition of the EMT is:
\begin{align}
   \label{gen-tmunudef}
 T_{\mu\nu}=-\frac{2}{\sqrt{-g}}\frac{\delta(\sqrt{-g}\mathcal{L}_{\rm m})}{\delta g^{\mu\nu}}=\mathcal{L}_{\rm m} g_{\mu\nu}-2\frac{\partial \mathcal{L}_{\rm m}}{\partial g^{\mu\nu}}.
 \end{align}
Consequently, the field equations are given by
\begin{align}  \label{modfieldeq}
G_{\mu\nu}+\Lambda g_{\mu\nu}=\kappa T_{\mu\nu}+\kappa T_{\mu\nu}^{{\rm{mod}} },
\end{align}
where $G_{\mu\nu}=R_{\mu\nu}-\frac{1}{2}Rg_{\mu\nu}$ is the Einstein tensor. Here $T_{\mu\nu}^{\rm{mod}}$ denotes the new terms coming from the metric variation of $\sqrt{-g} f$ as follows:
\begin{equation}   \label{def-mod}
T_{\mu\nu}^{\rm{mod}}=-\frac{2}{\sqrt{-g}}\frac{\delta(\sqrt{-g}f)}{\delta g^{\mu\nu}}=f g_{\mu\nu}-2 f_{\mathbf{T^2}} \Theta_{\mu\nu},
\end{equation}
where $f_{\mathbf{T^2}} \equiv \frac{\delta f}{\delta\mathbf{T^2}}$, and the new tensor $\Theta_{\mu\nu}\equiv \frac{\delta\, \mathbf{T^2}}{\delta g^{\mu\nu}}$, utilizing Eq.~\eqref{gen-tmunudef}, yields 
\begin{equation}
\begin{aligned}  
\label{thetadef}
\Theta_{\mu\nu}
&=-2\mathcal{L}_{\rm m}\left(T_{\mu\nu}-\frac{1}{2}g_{\mu\nu}\mathcal{T}\right)-\mathcal{T} T_{\mu\nu}\\
&\quad\,+2T_{\mu}^{\lambda}T_{\nu\lambda}-4T^{\sigma\epsilon}\frac{\partial^2 \mathcal{L}_{\rm m}}{\partial g^{\mu\nu} \partial g^{\sigma\epsilon}},
\end{aligned}
\end{equation}
where $\mathcal{T}=g_{\mu\nu} T^{\mu\nu}$ is the EMT's trace. Notably, all terms in the above equation, except the last one -- the second derivative of $\mathcal{L}_{\rm m}$ with respect to the inverse metric tensor -- are familiar matter-related scalars/tensors. This new term arises in gravity models containing the first metric derivative of $T_{\mu\nu}$. In the context of a perfect fluid, this term has been conventionally assumed to be zero in the literature, although this is not strictly accurate. However, even through reversing the deduction, substituting either $\mathcal{L}_{\rm m}=-\rho$ or $\mathcal{L}_{\rm m}=p$ into Eq.~\eqref{gen-tmunudef} for the EMT of perfect fluid unusually brings together the metric derivatives of $\rho$ (energy density) and $p$ (pressure) in this term~\cite{Brown:1993,Taub:1954,Schutz1970}. Moreover, properly calculating this term for the perfect fluid case for $\mathcal{L}_{\rm m}=p$ choice results in a divergence when $p=0$. For a comprehensive discussion within the thermodynamics framework, we refer readers to Ref.~\cite{Akarsu:2023lre}. This situation raises the question of how models that contain the second metric derivative of matter Lagrangian density studied in the literature to date have not suffered from any inconsistencies since this term is assumed to be zero though it is not indeed. Interestingly, this inquiry leads us to a more fundamental realization regarding EMSG as well as other matter-type modified gravity theories. It can be seen from redefining the material part of the Lagrangian density as $\mathcal{L}_{\rm m}^{\rm tot}=\mathcal{L}_{\rm m}+f$ that these models are equivalent to GR because the definition~\eqref{gen-tmunudef} is arranged in such a way that $T_{\mu\nu}$ emerges on the right-hand side of Einstein field equations. However, what differs is that these are nonminimally interacting two-fluid models in which the conventional EMT, $T_{\mu\nu}$, associated with the matter Lagrangian density $\mathcal{L}_{\rm m}$ is accompanied by a modification field $T_{\mu\nu}^{\rm mod}$. Imitating the definition stated in Eq.~\eqref{gen-tmunudef} gives rise to the total EMT as follows:
\begin{align} \label{emt-tot}
T_{\mu\nu}^{{\rm tot} }=-\frac{2}{\sqrt{-g}}\frac{\delta(\sqrt{-g}\mathcal{L}_{\rm m}^{{\rm tot} })}{\delta g^{\mu\nu}}=T_{\mu\nu}+T_{\mu\nu}^{{\rm mod} }.
\end{align}
From Eq.~\eqref{modfieldeq}, the twice contracted Bianchi identity, $\nabla^{\mu} G_{\mu\nu}=0$, leads to $\nabla^{\mu}(T_{\mu\nu}+T_{\mu\nu}^{{\rm mod} })=0$. However, this does not  necessarily imply $\nabla^{\mu}T_{\mu\nu}=0$ or $\nabla^{\mu} T_{\mu\nu}^{{\rm mod}}=0$. Thus, in such models, there is a potential nonminimal interaction between the material field and its counterpart, the modification field, which can be represented as:
\begin{equation}
\label{eq:int}
\nabla^{\mu}T_{\mu\nu}=-\mathcal{Q}_{\nu} \quad \textnormal{and} \quad \nabla^{\mu}T_{\mu\nu}^{{\rm mod} }=\mathcal{Q}_{\nu},
\end{equation}
where $\mathcal{Q}_{\nu}$ is the four-vector that functions as the interaction kernel, governing the energy transfer rate. The interaction kernel then assumes a specific form:
\begin{equation}
\label{eq:int1}
  \mathcal{Q}_{\nu}=\nabla_{\nu}f-2 \,\Theta_{\mu\nu} \nabla^{\mu} f_{\mathbf{T^2}} -2f_{\mathbf{T^2}}\nabla^{\mu} \Theta_{\mu\nu}. 
\end{equation}
This interaction model differs from other interaction models in that the function $f$ and the new tensor $\Theta_{\mu\nu}$  coexist in both Eqs.~\eqref{def-mod} and~\eqref{eq:int1}, intertwining the EoS of the second source described by $T_{\mu\nu}^{\rm mod}$ and the interaction kernel $\mathcal{Q}_{\nu}$, see Ref.~\cite{Akarsu:2023lre} for a detailed explanation. We opt to use only the function $f$, which dictates how the Lorentz scalar, ${\bf T^{2}}=T_{\mu \nu }T^{\mu \nu }$, is incorporated into the model, to formulate the interaction kernel. This choice, along with the undefined EoS of the source generated by $ T_{\mu\nu}^{\rm mod}$, provides flexibility in determining $\Theta_{\mu\nu}$. Thus, given the leeway in defining this tensor, to avoid concurrent emergence of the metric derivatives of $\rho$ and $p$, and accurately describe pressureless matter/dust with $\mathcal{L}_{\rm m}= p$, it is reasonable to omit the second metric derivative of $\mathcal{L}_{\rm m}$ in line with conventional literature. Consequently, we define the EMSF from Eq.~\eqref{def-mod} as follows:
\begin{align}
  \label{def-EMSF}
T_{\mu\nu}^{\rm EMSF}=f\,g_{\mu\nu}-2f_{\mathbf{T^2}}\Theta_{\mu\nu}^{\rm EMSF},
\end{align}
where the new tensor in the EMSF interaction model is given by:
\begin{align} \label{theta-EMSF}
\Theta_{\mu\nu}^{\rm EMSF}=-2\mathcal{L}_{\rm m}\left(T_{\mu\nu}-\frac{1}{2}g_{\mu\nu}\mathcal{T}\right) -\mathcal{T} T_{\mu\nu}+2T_{\mu}^{\lambda}T_{\nu\lambda}.  \end{align}

From Eq.~\eqref{eq:int1}, it is evident that even simple choices for the function $f$ can yield nontrivial interaction kernels with a covariant formulation.  It should be noted that with the aforementioned flexibility to exclude the last term from Eq.~\eqref{thetadef}, the models examined in the literature thus far have not encountered inconsistencies. However, this exclusion means that $\mathcal{L}_{\rm m}+f$ can no longer be considered as the total matter Lagrangian density of the EMSF interaction model, as it omits a term arising from the variation of $\sqrt{-g} f$ (see Eq.~\eqref{def-mod}). In other words, removing the second derivative of $\mathcal{L}_{\rm m}$ implies that the EMSF interaction lacks a well-defined Lagrangian formulation and operates effectively only at the field equation level. This case is valid for all the models -- even matter-curvature coupling theories such as $f(R_{\mu\nu}T^{\mu\nu})$~\cite{Haghani:2013oma,Odintsov:2013iba} and $f(G_{\mu\nu} T^{\mu\nu})$~\cite{Asimakis:2022jel} -- 
 in which this term is taken as zero. Before closing this discussion, we also note that all results presented in this section can be applied to other matter-type modified gravity theories by replacing $f({\bf T^{2}})$, for instance, with $f(\mathcal{L}_{\rm m})$~\cite{Harko:2010mv} or $f(\mathcal{T})$~\cite{Harko:2011kv}.

\section{\lowercase{q}EMSF extension of the standard cosmological model}
\label{sec:qEMSF}

In this paper, we investigate the early-time cosmological behavior of EMSF interaction model for the specific function $f({\bf T^{2}}) \propto {\bf T^{2}}$. We proceed by considering perfect fluids, described using the EMTs of the form:
\begin{align}
\label{em}
T_{\mu\nu}^{(i)}=(\rho_{i}+p_i)u_{\mu}u_{\nu}+p_i g_{\mu\nu},
\end{align} 
where $\rho_i$ and $p_i$ are the energy density and pressure of the $i$th fluid, respectively, and $u_{\mu}$ is the four-velocity of the medium, satisfying $u_{\mu}u^{\mu}=-1$. By choosing $\mathcal{L}_{\rm m}^{(i)}=p_i$, and integrating Eq.~\eqref{em} into Eq.~\eqref{theta-EMSF}, we derive
\begin{equation} 
\label{theta-pf}
\Theta_{\mu\nu}^{{\rm EMSF}(i)}=-(\rho_i+p_i) (\rho_i+3p_i) u_{\mu} u_{\nu}, 
\end{equation}
the standard expression used for the perfect fluids in existing EMSG models in the literature. We explore the most straightforward form of the EMSF, which we refer to as quadratic-EMSF (qEMSF), defined by
\begin{equation}
\label{function}
f({\bf T^{2}})=\sum_i \alpha_i T_{\mu\nu}^{(i)}T^{\mu\nu}_{(i)},
\end{equation}
where $\alpha_i$ is a parameter determining the extent of qEMSF relative to $T_{\mu\nu}^{(i)}$, and the summation over $i$ avoids cross-terms involving the EMTs of different sources in the field equations (thus, each $T_{\mu\nu}^{(i)}$ is accompanied by its own qEMSF), as detailed in Ref.~\cite{Akarsu:2018aro}. We also assume different material fields interact minimally, i.e., only gravitationally with each other. From Eq.~\eqref{def-EMSF} along with Eq.~\eqref{theta-pf}, when perfect fluid serves as the usual material field, the EMT of the accompanying qEMSF takes the form
 \begin{equation}
 \begin{aligned}
 T_{\mu\nu}^{{\rm qEMSF} (i)}= \alpha_i\left[(\rho_i^2+3p_i^2)(g_{\mu\nu}+2 u_{\mu} u_{\nu})+8\rho_i p_i u_{\mu} u_{\nu}\right].
 \end{aligned}
 \end{equation}
Next, we assume a spatially maximally symmetric spacetime metric (viz., the RW metric) with flat space-like sections,
\begin{align}
\label{RW}
{\rm d}s^2=-{\rm d}t^2+a^2\,({\rm d}x^2+{\rm d}y^2+{\rm d}z^2),
\end{align}  
where $a=a(t)$ is a function of cosmic time $t$ only, and also assume that the perfect fluid is comoving having the four-velocity $u^{\mu}=\delta^{\mu}_0$. By combining the influences of both the usual material field and qEMSF as per Eq.~\eqref{modfieldeq}, we arrive at
 \begin{align}
  &3 H^2=\Lambda+\kappa \sum_{i}\left[\rho_i+\alpha_i(\rho_i^2+8\rho_i p_i+3p_i^2)\right], \\
  &-2\dot{H}-3H^2=-\Lambda+\kappa \sum_{i} \left[p_i+\alpha_i(\rho_i^2+3p_i^2)\right],
 \end{align}
where $H=\frac{\dot{a}}{a}$ is the Hubble parameter, with an overdot indicating derivative with respect to the cosmic time $t$.

Utilizing a barotropic fluid with EoS parameter  $w_i=p_i/\rho_i={\rm const.}$, we assume, as is customary, that the universe contains pressureless matter/dust with an energy density $\rho_{\rm m}$ and an EoS parameter $w_{\rm m}=0$, along with radiation/relativistic fluid having an energy density $\rho_{\rm r}$ and an EoS parameter $w_{\rm r}=\frac{1}{3}$. In line with Occam’s razor, we opt for $\alpha\equiv\alpha_{\rm r}=\alpha_{\rm m}$, thereby reducing the number of free parameters in our model.\footnote{Since we confine the discussion in this work to the BBN epoch, it is sufficient for us to focus on the radiation-dominated epoch of the Universe. For this reason, we assume $\alpha_i$'s are same for each component inspired by Occam's razor, viz. $\alpha=\alpha_{\rm m}=\alpha_{\rm r}$. An alternative approach would be taking $\alpha_{\rm m}=0$ meaning that matter does not have a qEMSF partner. Nevertheless, our point here is merely to reduce the number of free parameters, and in both cases, matter contributions will be negligible during the cosmological era that concerns us. On the other hand, allowing $\alpha_{i}$ parameters differ for each species may provide us with the opportunity to enhance the model by modifying both the early and late times of the Universe in order to address, for instance, $H_0$ tension. This approach may have far-reaching implications on the model and pave way for interesting novel studies; for instance, see Refs.~\cite{Akarsu:2018aro,Akarsu:2020vii}.}
Consequently, we arrive at the following set of two linearly independent differential equations with three unknown functions $H$, $\rho_{\rm r}$, and $\rho_{\rm m}$: 
\begin{align}
3 H^2=\Lambda+\kappa\left(\rho_{\rm r}+4\alpha\rho_{\rm r}^2\right)+\kappa\left(\rho_{\rm m}+\alpha \rho_{\rm m}^2\right),\label{eq:rhoprime1}\\
-2\dot{H}-3H^2=-\Lambda+\frac{\kappa}{3}\left(\rho_{\rm r}+4\alpha\rho_{\rm r}^2\right)+\kappa\alpha \rho_{\rm m}^2. 
 \label{eq:presprime1}
\end{align}
 The energy density associated with the $\Lambda$, well constrained at $\rho_{\Lambda}=\Lambda/\kappa \sim 10^{-12}\,{\rm eV^{4}}$ by cosmological observations and comparable to the present-day critical energy density of the Universe~\cite{Planck15Cosmo}, is negligible compared to the energy density scales of BBN ($\rho_{\rm bbn} \sim 10^{22}\,{\rm eV^{4}}$~\cite{Dodelson03}) and neutron stars ($\rho_{\rm ns} \sim 10^{34}\,{\rm eV^{4}}$~\cite{Shapiro83}). The qEMSF-related terms become significant if $\rho \sim |\alpha|\rho^2$, suggesting that qEMSF interaction effects would be observable in the present-day Universe dynamics for $|\alpha|\sim 10^{12}\,{\rm eV^{-4}}$, and during the BBN processes if $|\alpha| \sim 10^{-22}\,{\rm eV^{-4}}$. Consequently, within the qEMSF interaction framework, we preliminarily restrict  $|\alpha| \lesssim 10^{-22}\,{\rm eV^{-4}}$ to ensure minimal deviation from standard dynamics of BBN. 

As a further step, we examine the deceleration parameter, defined as $q \equiv -1-\frac{\dot H}{H^2}$. For the qEMSF interaction, this parameter is given by
\begin{align}
\label{decparam1}
q=-1+2\frac{\left(1+4\alpha\rho_{\rm r}\right)\rho_{\rm r}+\left(\frac{3}{4}+\frac{3\alpha}{2} \rho_{\rm m}\right)\rho_{\rm m}}{\frac{\Lambda}{\kappa}+\left(1+4\alpha\rho_{\rm r}\right)\rho_{\rm r}+\left(1+\alpha \rho_{\rm m}\right)\rho_{\rm m}},
\end{align}
which is utilized to assess potential deviations from the standard cosmology (minimal interaction case). At early epochs, specifically during the BBN epoch when radiation is dominant -- and disregarding pressureless matter and $\Lambda$ in Eq.~\eqref{decparam1} -- the deceleration parameter in the qEMSF interaction model aligns with that of the standard radiation-dominated universe. That is,
\begin{align}
q_{\rm bbn,\Lambda\rm{CDM}}=q_{\rm bbn, qEMSF}=1,
\end{align}
resulting in a scale factor $a$ proportional to $t^{1/2}$ in both standard and qEMSF interaction models.

To close the system consisting of Eqs.~\eqref{eq:rhoprime1}-\eqref{eq:presprime1}, we adhere to our previous assumption that all fluids -- matter and radiation -- along with their qEMSF counterparts, interact only gravitationally. This leads to the separation of the continuity equation into two distinct parts:
\begin{align}
\label{rhomsol1}
\dot \rho_{\rm m}+ 3H\rho_{\rm m}&=0,\\
  \dot \rho_{\rm r}+ 4H\frac{1+4\alpha\rho_{\rm r}}{1+8\alpha\rho_{{\rm r}}}\rho_{\rm r}&=0,
\label{rhorsol1}
\end{align}
which respectively yield the following scale factor dependencies for matter and radiation:
\begin{align}
\label{rhomsol}
\rho_{\rm m}&=\rho_{\rm m0} a^{-3},\\ 
\rho_{\rm r}&=\frac{1}{8\alpha}\left(\sqrt{1+16\alpha\rho_{\rm r0}(1+4\alpha\rho_{\rm r0})a^{-4}}-1\right),
\label{rhorsol}
\end{align}
where $\rho_{\rm m0}>0$ and $\rho_{\rm r0}>0$ are the present-day energy densities of dust and radiation, respectively -- here and hereafter, a subscript $0$ attached to any quantity denotes its present-day (i.e., at $a=1$) value. Note also that the solution referred to as Eq.~\eqref{rhorsol} is valid provided that the condition $\alpha>-\frac{1}{8 \rho_{\rm r0}}$ is satisfied. It can be verified that under this condition, $\rho_{\rm r}=\rho_{\rm r0}$ when $a=1$, and it also ensures that $\rho_{\rm r}$ remains always positive. Finally, we can demonstrate that $\rho_{\rm r}\rightarrow\rho_{\rm r0}a^{-4}$ as $\alpha\rightarrow0$, ensuring that the radiation-dominated epoch of the standard model is accurately recovered. In the qEMSF interaction model, employing the new tensor~\eqref{theta-EMSF} of EMSF along with the function $f(\mathbf{T^2})=\alpha \mathbf{T^2}$ in the interaction kernel~\eqref{eq:int}, we obtain 
\begin{align}
Q^{\rm (m)}_0=0& \qquad \textnormal{(for matter)}, \\ Q^{\rm (r)}_0=8 \alpha \dot{\rho}_{\rm r} \rho_{\rm r}+16 \alpha H \rho_{\rm r}^2& \qquad \textnormal{(for radiation)},
\end{align}
where the latter can be expressed as $Q^{\rm (r)}_0=-\frac{16 \alpha \rho_{\rm r}^2}{1+8 \alpha \rho_{\rm r}}H$ when eliminating $\dot{\rho}_{\rm r}$ through Eq.~\eqref{rhorsol1}. We would like to state that these kernels can be directly derived from the continuity equations given in Eqs.~\eqref{rhomsol1} and~\eqref{rhorsol1} as well.

Since pressureless matter in our model behaves as in the standard cosmology, its qEMSF partner's energy density, the last term in Eq.~\eqref{eq:rhoprime1}, evolving with $a^{-6}$, mimics the stiff fluid contribution to the Friedmann equation as exactly like the shear scalar that retains isotropic spatial curvature does too.\footnote{For further insights, see Ref.~\cite{Chavanis:2014lra}, which examines cosmological models incorporating a stiff fluid source atop of the $\Lambda$CDM model, particularly regarding the evolution of the expansion scale factor in our model, and Ref.~\cite{Barrow78} for an investigation of anisotropic cosmologies in the presence of stiff fluid with a positive energy density.} The observational upper limits on the present-day density parameter of a stiff fluid-like term, as suggested by Refs.~\cite{Akarsu:2019pwn,Akarsu:2020vii,Akarsu:2021max}, are $\Omega_{\rm s0} \sim 10^{-15}$ from the latest cosmological data (viz., joint CMB and BAO dataset), and $\Omega_{\rm s0} \sim 10^{-23}$ from BBN.\footnote{If the constraint from the joint CMB and BAO dataset, $\Omega_{\rm s0} \sim 10^{-15}$, is considered, this term enhances the Hubble expansion rate in the post-recombination universe, thereby similarly increasing $H_0$, by reducing the sound horizon in a similar fashion to Early Dark Energy (EDE) models~\cite{Poulin:2018cxd}. One might expect the models allowing anisotropic expansion on top of $\Lambda$CDM to alleviate the Hubble tension~\cite{Akarsu:2019pwn,Akarsu:2021max,DiValentino:2021izs}, yet expansion anisotropy becomes irrelevant to make a significant change in the $H_0$ value when the tight BBN constraint, $\Omega_{\rm s0} \sim 10^{-23}$, is imposed. Namely, it is not possible to resolve the $H_0$ tension without spoiling the successes of the standard BBN. Nevertheless, one can still control the behavior of the shear scalar by introducing anisotropic source or qEMSF interaction alongside anisotropic expansion~\cite{Akarsu:2020vii,Barrow:1997sy,Akarsu:2020pka}.} The relation $\Omega_{\rm s0}\sim \alpha\rho_{\rm m0}^2/\rho_{\rm cr0}$, where the critical energy density is defined as $\rho_{\rm cr}=3H^2/\kappa$, translates the constraint on $\Omega_{\rm s0}$ to $|\alpha| \sim 2.9\times 10^{-12}\,{\rm eV^{-4}}$, using $\rho_{\rm cr0}=8.1\times 10^{-11} h^2\, {\rm eV^{4}}$,  $\Omega_{\rm m0}=0.14\,h^{-2}$~\cite{Aghanim}, with $\Omega_{\rm m0}=\rho_{\rm m0}/\rho_{\rm cr0}$ being the present-day density parameter of matter and $h=0.68$ with $h=H_0/100\,{\rm km\,s}^{-1}{\rm Mpc}^{-1}$ as the dimensionless reduced Hubble constant. Comparing the preliminary constraint $|\alpha| \lesssim 10^{-22}\,{\rm eV^{-4}}$ from BBN with the constraint on the stiff fluid-like term, the quadratic contribution of pressureless matter can be considered negligible, and BBN remains an epoch dominated solely by radiation in the qEMSF interaction model. In this model, energy-momentum conservation is violated for radiation, leading to an evolution of its energy density that differs from $\rho_{\rm r} \propto a^{-4}$ -- due to the multiplier of $H$ in Eq.~\eqref{rhorsol1} -- while the deceleration parameter $q_{\rm bbn}=1$ and the standard matter evolution in the radiation-dominated epoch remain unchanged. This raises intriguing questions, such as the implications of the thermodynamic analysis of usual radiation separated from its qEMSF counterpart with which it interacts nonminimally, which we will examine in the following section.

\section{Thermodynamics of the model}
\label{sec:thermo}

Nonvanishing divergence of the usual EMT -- resulting from nonminimal matter interactions, as in our model, and also in modified theories of gravity with nonminimal geometry-matter couplings -- can be examined from a thermodynamic viewpoint. Although radiation and its accompanying qEMSF counterpart together form a closed system, analyzing the energy-momentum transfer between them requires applying thermodynamics of open systems to each component separately. This approach enables the investigation of the relationship between the direction of energy-momentum transfer and entropy changes in each component, particularly focusing on the influence of the parameter $\alpha$.

Extending the concept of an adiabatic process 
(characterized by no heat transfer, ${\rm d}Q=0$) from closed to open systems, as first presented in Ref.~\cite{Prigogine:1989zz}, i.e., systems containing $N$ particles, which may not be constant in number within a volume element $V=a^3$, yields the thermodynamic energy conservation law:
\begin{align}
\label{eq:thermo}
{\rm d}(\rho V)+p{\rm d}V-\frac{\varrho}{n}{\rm d}(nV)=0,    
\end{align}
where $n=N/V$ is the particle number density, and $\varrho=\rho+p$ is the enthalpy per unit volume (or inertial mass density). In contrast to closed systems, assumed in the standard cosmology, where entropy change ${\rm d}S$ vanishes in adiabatic processes, in open systems, it is given by:
\begin{align}
\label{eq:thermoS}
T{\rm d}S=\left(\frac{\varrho}{n}-\mu\right) {\rm d}(nV)=\frac{Ts}{n}{\rm d}(nV),    
\end{align}
with $\mu n=\varrho-T s$ being the chemical potential and $s=S/V$ as the entropy density. Combining Eqs.~\eqref{eq:thermo} and~\eqref{eq:thermoS}, we obtain:
\begin{align}
\label{eq:def}
\dot{\rho}+3H\varrho=\varrho \left[\frac{\dot{n}}{n}+3H\right]=\varrho \frac{\dot{S}}{S} .
\end{align}
In our model, both pressureless matter and $\Lambda$ evolve as in the standard cosmology. However, radiation, with inertial mass density $\varrho_{\rm r}=\frac{4}{3}\rho_{\rm r}$, contributes to potential entropy changes due to its nonconservation. Specifically, we have
\begin{align}
\label{eq:def1}
\dot{\rho}_{\rm r}+4H\rho_{\rm r}=\frac{4}{3}\rho_{\rm r}\left[\frac{\dot{n}}{n}+3H\right], 
\end{align}
resulting in:
\begin{align}  \label{entropy}
\frac{3}{4}\frac{\dot{\rho}_{\rm r}}{\rho_{\rm r}}=\frac{\dot{n}}{n}\quad\rightarrow\quad -\frac{9}{4}H\left(1+w_{\rm r,tot}\right)=\frac{\dot{S}}{S}-3H,
 \end{align}
where Eq.~\eqref{rhorsol1} is recast as $\dot{\rho}_{\rm r}+3H\rho_{\rm r}(1+w_{\rm r,tot})=0$, indicating that radiation along with its qEMSF partner, contributes to the Friedmann equation as a source with an EoS parameter of $w_{\rm r,tot}$.

\begin{figure}[t!]
  \includegraphics[width=0.48\textwidth]{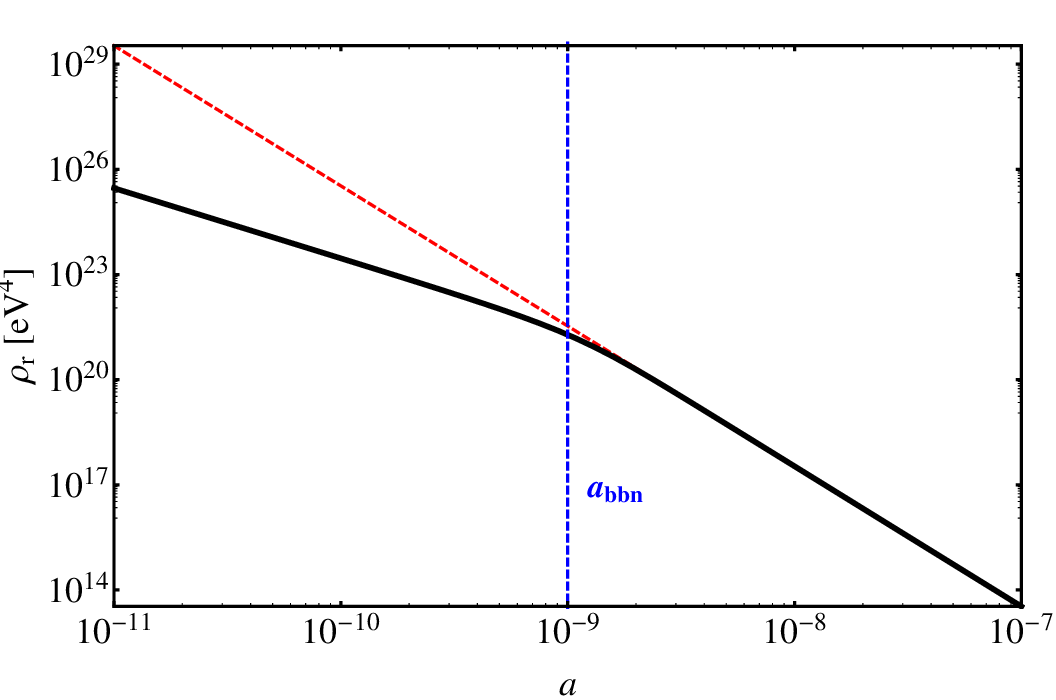}
  \caption{$\rho_{\rm r}$ versus $a$ in the standard cosmological model (dotted red line) and the qEMSF interaction model (black line), with $\alpha=10^{-22}\,{\rm eV^{-4}}$ and $\rho_{\rm r0}=3.38 \times 10^{-15}\,{\rm eV^{4}}$ (present-day energy density of radiation). The vertical line (blue) represents the BBN era.}
  \label{fig:rhorad}
\end{figure}

Employing the entropy change relation as stated in Eq.~\eqref{entropy}, we deduce:
\begin{align}   \label{2ndlaw}
\frac{\dot{S}}{S}=\frac{3}{4}H(1-3w_{\rm r,tot}).
 \end{align}
In an expanding universe ($H>0$), this equation shows that the entropy of the usual radiation component increases when $w_{\rm r,tot}<\frac{1}{3}$, decreases for $w_{\rm r,tot}>\frac{1}{3}$, and remains constant for $w_{\rm r,tot}=\frac{1}{3}$, corresponding to the $\Lambda$CDM case ($\alpha \rightarrow 0$ limit) without qEMSF partner. Utilizing Eq.~\eqref{rhorsol}, the total EoS parameter is determined as:
 \begin{align}
\label{eq:wreff}
w_{\rm r,tot}=\frac{1}{3}\left[-1+\frac{2}{\sqrt{1+16\alpha\rho_{\rm r0}(1+4\alpha\rho_{\rm r0})a^{-4}}}\right].
\end{align}
We observe that under the condition $\alpha>-\frac{1}{8 \rho_{\rm r0}}$ coming along with the solution~\eqref{rhorsol}, $w_{\rm r,tot}<\frac{1}{3}$ for $\alpha>0$ and $w_{\rm r,tot}>\frac{1}{3}$ for $\alpha<0$. Thus, Eqs.~\eqref{2ndlaw} and~\eqref{eq:wreff} suggest that positive values of $\alpha$ lead to an increase in the entropy of the usual radiation component, and conversely, negative values lead to a decrease. This implies a corresponding decrease or increase in the entropy of its qEMSF counterpart, as the total entropy of a closed system (radiation+its qEMSF counterpart) remains constant in an adiabatic process.  Additionally, Eq.~\eqref{eq:wreff} shows that for $\alpha>0$, the total EoS parameter ranges between $-\frac{1}{3}\leq w_{\rm r,tot}< \frac{1}{3}$, suggesting that, according to the continuity equation, $\frac{{\rm d} \ln  \rho_{\rm r}}{{\rm d} \ln a}=-3(1+w_{\rm r,tot})$,
radiation can dilute at different rates, such that it can be the slowest diluting component with $\rho_{\rm r}\propto a^{-2}$ and also the fastest diluting component with almost $\rho_{\rm r}\propto a^{-4}$ (i.e., nearly as in the standard cosmology, without qEMSF partner).  This indicates that a positive $\alpha$ leads to a slower dilution of radiation energy density compared to the standard cosmology; the usual radiation component transfers energy-momentum from its qEMSF counterpart, where the maximum transfer leads that radiation mimics the role of positive spatial curvature. Fig.~\ref{fig:rhorad} illustrates the evolution of radiation energy density with respect to the scale factor, considering $\alpha=10^{-22}\,{\rm eV^{-4}}$ and $\rho_{\rm r0}=3.38 \times 10^{-15}\,{\rm eV^{4}}$ (present-day energy density of radiation). The behavior of radiation is observed to approximate that of the standard cosmology for $a\gtrsim  a_{\rm bbn}$, with slight deviation at the BBN epoch ($a_{\rm bbn} \sim 10^{-9}$), and exhibits a considerable deviation in pre-BBN times. As our discussion is confined to the times relevant to BBN, we refer the reader to Ref.~\cite{Akarsu:2018zxl} 
for a detailed exploration of pre-BBN dynamics, in particular, to Fig.~7b of this work for the $\rho_{\rm r}$ behavior in cases of different positive/negative $\alpha$ values. In nonminimal matter interaction models, energy-momentum transfer between the sources does not necessarily prefer a specific direction.\footnote{In contrast, nonminimal matter-curvature coupling theories exhibit an asymmetrical relationship between spacetime geometry and material stresses. This asymmetry arises because particle creation from gravitational energy is feasible, whereas the reverse process (annihilation) is not. Consequently, in these theories, the particle production criterion ${\rm d}(nV)\geq 0$ as derived from Eq.~\eqref{eq:thermo} necessitates that the entropy change ${\rm d}S$ for the material component must be nonnegative (${\rm d}S\geq 0$).}\footnote{In earlier interpretations of EMSG, as well as other matter-type gravity models considered in the literature, the usual material field is posited to be nonminimally coupled to spacetime curvature. This perspective dictates that energy flow is exclusively directed from the gravitational field to the usual material component, effectively prohibiting any decrease in particle number. As such, this interpretation specifically imposes that ${\rm d}S\geq 0$. Consequently, within the EMSG framework, adopting this approach necessitates restricting the $\alpha$ parameter to nonnegative values ($\alpha \geq 0$). This condition is emphasized for readers interested in examining our model from this viewpoint. However, for a critical analysis of this interpretation regarding matter-type models and for further discussion, we refer readers to Ref.~\cite{Akarsu:2023lre}.} In our model, when $\alpha<0$, we have $w_{\rm r,tot}>\frac{1}{3}$, implying that the usual radiation dilutes faster than $\rho_{\rm r} \propto a^{-4}$ and consequently loses entropy and nurtures its qEMSF partner. These can also be verified from Eq.~\eqref{rhorsol1}, showing that in an expanding universe, $\dot \rho_{\rm r}+ 4H \rho_{\rm r} > 0$ for $\alpha > 0$ (indicating radiation gains energy-momentum), and $\dot \rho_{\rm r}+ 4H \rho_{\rm r} < 0$ for $\alpha<0$ (indicating radiation loses energy-momentum). Having determined the direction of energy-momentum/entropy transfer depending on $\alpha$, the modified evolution of radiation, as given in Eq.~\eqref{rhorsol}, while preserving both $q_{\rm bbn}=1$ and the scale factor dependence of matter, prompts us to investigate BBN era, during which radiation dominates, in the framework of qEMSF interaction model. Thus, in what follows we study the constraints on $\alpha$ from BBN epoch and then discuss the implications of the results on the BBN physics.

\section{BBN constraints on the model}
\label{sec:bbn}
At the time of BBN, corresponding to the scale factor $a_{\rm bbn}\simeq10^{-9}$, the ratio of radiation to matter density can be expressed from Eqs.~\eqref{rhomsol} and~\eqref{rhorsol}, taking into account that $a_{\rm bbn} \ll 1 $, as:
 \begin{align} \label{rad-dom}
\bigg[\frac{\rho_{\rm r}}{\rho_{\rm m}}\bigg]_{\rm bbn} \simeq \frac{\Omega_{\rm r0}}
{\Omega_{\rm m0}}\frac{\sqrt{a_{\rm bbn}^4+16\alpha\rho_{\rm r0} }a_{\rm bbn}-a_{\rm bbn}^{3}}{8\alpha\rho_{\rm r0}}\gtrsim1,
\end{align}
indicating that radiation dominates over matter if the condition
\begin{align}
   \label{eq:alphaconst}
 -1.8\times 10^{-23}\,{\rm eV^{-4}}\lesssim \alpha \lesssim 6.6\times10^{-12} \,{\rm eV^{-4}}
\end{align}
is satisfied. Here, the present-day density parameter of radiation is defined as $\Omega_{\rm r0}= \rho_{\rm r0}/\rho_{\rm cr0}$ and since
the present-day photon energy density is extremely well constrained by the absolute CMB monopole temperature measured by FIRAS $T_{\rm CMB} = 2.7255 \pm 0.0006 {\rm K}$~\cite{Fixsen:2009ug}, we have used
$\Omega_{\rm r0}=2.469\times10^{-5}h^{-2}(1+0.2271N_{\nu,{\rm eff}})$~\cite{WMAP:2010qai} with $N_{\nu,{\rm eff}}=3.046$, the standard number of effective neutrino species with minimum allowed mass $m_{\nu} = 0.06\,{\rm eV}$. Compared to the preliminary constraint previously discussed ($-10^{-22} \lesssim \alpha \lesssim 10^{-22}\,{\rm eV^{-4}}$), the constraint in Eq.~\eqref{eq:alphaconst} appears less stringent for positive $\alpha$ values but provides a slightly tighter bound for negative $\alpha$ values; refining our understanding of the permissible range of $\alpha$ in the context of early Universe cosmology.

\subsection{Early universe}
\label{sec:earlyuniverse}

Considering the constraint given in Eq.~\eqref{eq:alphaconst}, we can assume that in the early Universe, radiation (including photons and neutrinos) is dominant, while contributions from matter, spatial curvature (if present), and the cosmological constant are negligible. Accordingly, Eqs.~\eqref{eq:rhoprime1}-\eqref{eq:presprime1} simplify to:
\begin{align}   
\label{eq:rhoprime}
3H^2=&\kappa\rho_{\rm r}(1+4\alpha\rho_{\rm r}),\\
-2\dot{H}-3H^2=&\kappa \frac{\rho_{\rm r}}{3}(1+4\alpha\rho_{\rm r}).
\label{eq:presprime}
\end{align}
We note that the positivity of Eq.~\eqref{eq:rhoprime} at the BBN epoch necessitates $\alpha>-\frac{1}{4\rho_{\rm r,bbn}}$.  This condition, being significantly stricter due to $\rho_{\rm r,bbn}\gg\rho_{\rm r0}$, automatically ensures the earlier $\alpha>-\frac{1}{8 \rho_{\rm r0}}$ condition associated with the solution given Eq.~\eqref{rhorsol}. The realistic solution of this system, Eqs.~\eqref{eq:rhoprime}-\eqref{eq:presprime}, is given as follows:
\begin{equation}
\label{scalerho}
a=a_1\,t^{1/2}\quad\textnormal{and}\quad\rho_{\rm r}=\frac{1}{8\alpha}\left(\sqrt{1+\frac{12\alpha}{\kappa t^2}}-1\right),
\end{equation}
where the behavior of $\rho_{\rm r}$ corresponds to the one given in Eq.~\eqref{rhorsol}, as discussed in Ref.~\cite{Akarsu:2018zxl}.\footnote{We refer the reader to Ref.~\cite{Board:2017ign} for a comprehensive analysis of realistic cosmological solutions in qEMSF interaction model. However, in relevance with the discussion we carry out here, it would be useful to mention the other solution of the system~\eqref{eq:rhoprime}-\eqref{eq:presprime} which is $a=a_1\,t^{1/2}$ and $\rho_{\rm r}=-\frac{1}{8\alpha}\left(1+\sqrt{1+\frac{12\alpha}{\kappa t^2}}\right)$ as commented in Ref.~\cite{Akarsu:2018zxl}. We note however that, for this solution used in Ref.~\cite{Roshan:2016mbt}, $\rho_{\rm r}>0$ only if $\alpha<0$ and more importantly $\lim_{\alpha\to0^{-}}\rho_{\rm r}=\infty$ rather than approaching $\frac{3}{4\kappa t^2}$ that standard cosmology would give. This implies that this second solution does not recover the standard model limit properly.} Here, $a_1$ represents the cosmic scale factor length when the Universe is one second old (i.e., $t=1\,{\rm s}$). One may verify that in this solution, as $\alpha\rightarrow 0$, $\rho_{\rm r}\rightarrow\frac{3}{4\kappa t^2}$, properly recovering the radiation-dominated era of the standard cosmology. When $\alpha\neq0$, indicating that radiation is nonminimally interacting with its qEMSF partner, the time evolution of its energy density deviates from that in standard cosmology. For $\alpha\geq0$, we observe that as $t\rightarrow+\infty$, the scale factor diverges ($a\rightarrow+\infty$) and the radiation energy density approaches zero ($\rho_{\rm r}\rightarrow0$). Conversely, as $t\rightarrow 0$ (approaching big bang), $a$ approaches zero ($a\rightarrow0$) and $\rho_{\rm r}$ diverges ($\rho_{\rm r}\rightarrow+\infty$). For $\alpha<0$, the behavior is similar to  the $\alpha>0$ case in terms of the limit as $a\rightarrow+\infty$. However, when we look at the earliest times of the Universe, the radiation energy density reaches a finite maximum value $\left(\rho_{\rm r,max}=-\frac{1}{8\alpha}\right)$ when the cosmic scale factor is at its minimum $\left(a_{\rm min}=a_1\left(-\frac{12\alpha}{\kappa}\right)^{1/4}\right)$. It is noteworthy that the rate of change of $\rho_{\rm r}$, $\dot{\rho}_{\rm r}=-\frac{3}{2\kappa}\left(1+\frac{12\alpha}{\kappa t^2}\right)^{-1/2} \,t^{-3}$, is consistently negative, irrespective of the sign of $\alpha$. This indicates that the radiation energy density decreases monotonically with time. Accordingly, we anticipate that the deviation from the standard radiation-dominated universe will become significant at times earlier than certain time scales determined by the value of the parameter $\alpha$.

The time evolution of the Hubble parameter and the value of the deceleration parameter remain the same as in the standard radiation-dominated universe, that is, $H(t)=H_{\rm std}(t)=\frac{1}{2t}$ and $q=q_{\rm std}=1$. However, as can be seen from Eqs.~\eqref{eq:rhoprime}-\eqref{eq:presprime}, $H(\rho_{\rm r})$, the value of the Hubble parameter $H$ for a given radiation energy density value, differs from that in the standard model, $H_{\rm std}(\rho_{\rm r})=\sqrt{\frac{\kappa}{3}\rho_{\rm r}}\,$, as shown below:
\begin{equation}
\label{eq:Hubblemod}
\frac{H(\rho_{\rm r})}{H_{\rm std}(\rho_{\rm r})}=\sqrt{1+4\alpha \rho_{\rm r}}\,,
\end{equation}
where the altered scale factor dependency of $\rho_{\rm r}$ due to its nonconservation is given in Eq.~\eqref{rhorsol}. The presence of the qEMSF interaction at temperatures around MeV scale would lead to modifications in the Hubble rate and the time evolution of the temperatures of photons and neutrinos, which, in turn, would lead to changes in the predicted abundances of light nuclei. If we ensure that the energy density of the quadratic correction is of the same order of magnitude as the usual radiation term at the BBN epoch, we can deduce from Eq.~\eqref{eq:Hubblemod} that $|\alpha|\leq (4\rho_{\rm r,bbn})^{-1}$. Using $\rho_{\rm r,bbn} \sim 10^{22}\, {\rm eV^{4}}$~\cite{Dodelson03}, this approach allows us to refine the preliminary allowed range for the $\alpha$ parameter to:
\begin{equation}
-10^{-23}\,{\rm eV^{-4}}\lesssim \alpha \lesssim 10^{-23} \,{\rm eV^{-4}}.
\label{eq:preliconst}
\end{equation}
Furthermore, since the abundances of light elements are sensitive to the Universe's expansion rate during BBN epoch, the differences in cosmic history stemming from the altered dynamics of radiation and their potential impact on these abundances could provide a more stringent constraint on the $\alpha$ parameter.

\subsection{Primordial Nucleosynthesis}
\label{sec:prinuc}

Given that radiation maintains its blackbody characteristics in the early Universe, its energy density adheres to the standard temperature evolution, expressed as $\rho_{\rm r} = \frac{\pi^2}{30}g_* T^4$, where $g_*$ represents the effective number of relativistic degrees of freedom. At the BBN-relevant energy scales of around $1$ MeV, considering the contributions from photons ($\gamma$), electrons (${\rm e}^{-}$) and positrons (${\rm e}^{+}$), and neutrino flavors ($\nu$), we deduce $g_*= 2 + \frac{7}{2} + \frac{7}{4}N_\nu$ with $N_{\nu}=3$, which results in an effective neutrino number of $N_{\nu,\rm eff}=3.046$ within the Standard Model of particle physics. This effective number accounts for the increased number of neutrinos present in the thermal bath, owing to the elevated temperatures from ${\rm e}^+{\rm e}^-$ annihilations before complete neutrino decoupling. For a comprehensive understanding of the BBN formalism, refer to Refs.~\cite{bbf,kolb,Olive:1999ij,Cyburt:2015mya}. We observe from Eq.~\eqref{eq:Hubblemod} that the expansion rate of the Universe during the BBN epoch is altered due to the extra term introduced by the qEMSF interaction model. By utilizing the standard blackbody energy density for relativistic species, represented as $\rho_{\rm r}=3H_{\rm std}^2/\kappa$, this modification can be expressed as follows:
\begin{align}
\label{eq:modH}
H^2=&H^2_{\rm std}\left(1+\frac{12\alpha}{\kappa}H^2_{\rm std}\right),
\end{align}
where $H^2_{\rm std}= \frac{\kappa}{3} \frac{\pi^2}{30}g_* T^4$ is the Friedmann equation for the BBN epoch within the standard cosmological model (i.e., when $\alpha \rightarrow 0$, thus without qEMSF partner). This scenario is akin to modifying the effective number of degrees of freedom leading to $3 H^2_{\rm std}=\kappa \frac{\pi^2}{30} \tilde{g}_* T^4\,$, for instance, by introducing an extra relativistic species such as sterile neutrino.

At lower temperatures, when the rate of interactions falls behind the rate of expansion (i.e., $\Gamma(T)\ll H$), particles decouple at freeze-out temperature $T_{\rm f}$ as they lack the necessary time interval to interact. The freeze-out condition is determined by $H \approx \Gamma$, where $\Gamma=n \sigma v$ represents the relevant weak interaction rates per baryon. Here $n$ is the number density of particles, $\sigma$ is the interaction cross section, and $v$ is the average speed of the particles. The cross-section associated with the weak interaction scales as $\sigma \sim G_{\rm F}^2 T_{\rm f}^2$, where $G_{\rm F}=1.166\times10^{-5}\,{\rm GeV}^{-2}$ is the Fermi coupling constant (see Ref.~\cite{bbf} for details). In standard model, the scale factor's evolution as $a \propto 1/T$ during the radiation-dominated era leads to the particle number density evolving as $n \propto T^3$. Consequently, the rough scaling for the weak interaction rate becomes $\Gamma \sim G_{\rm F}^2 T_{\rm f}^5$. Additionally, factoring in the first-order correction proportional to $T^5$ in this scaling, the weak interaction rate can be approximated as~\cite{Torres:1997sn} 
\begin{align}   \label{eq:fo}
H(T_{\rm f}) \approx \Gamma(T_{\rm f}) \approx q\, T_{\rm f}^5,
\end{align}
where $q\simeq9.8\times10^{-10}\,{\rm GeV}^{-4}$~\cite{Torres:1997sn,Lambiase:2005kb,Lambiase:2011zz,Capozziello:2017bxm,Barrow:2020kug}. This equation indicates that the freeze-out temperature $T_{\rm f}$ is influenced by both gravitational and weak interactions. For a detailed analytical expression of the conversion rates of protons into neutrons and its inverse, see Ref.~\cite{bbf}.

However, the qEMSF interaction model introduces modifications to the scale factor-temperature relation during the radiation-dominated era, as shown by:
\begin{equation} \label{scale-T}
 a^4 = \frac{a_1^4}{\frac{4}{3}\kappa \frac{\pi^2}{30}g_* T^4 \left(1+\alpha \frac{4 \pi^2}{30} g_*  T^4\right)},
 \end{equation}
which also impacts the number density of particles. To maintain the validity of the weak interaction rate scaling from standard cosmology, as expressed in Eq.~\eqref{eq:fo}, the correction term involving $\alpha$ in Eq.~\eqref{scale-T} should be of the order of $\mathcal{O}(1)$ for temperatures pertinent to BBN ($T\sim 1\, \rm MeV$). Consequently, this requirement leads us to a preliminary, but stringent, constraint on $\alpha$: \begin{equation}
-10^{-25} \,{\rm eV^{-4}}\lesssim \alpha \lesssim 10^{-25} \,{\rm eV^{-4}}.
\label{eq:preliconst2}
\end{equation}

Now, applying Eq.~\eqref{eq:fo} to Eq.~\eqref{eq:modH}, we derive a relation for the freeze-out temperature of ${\rm \beta}$-interactions, which govern the relative abundances of neutrons and protons: 
\begin{align}
\label{eq:modT}
T_{\rm f}^{6}= T_{\rm fstd}^{6}\left(1+ 12\,\alpha\,\frac{q^2}{\kappa}\, T^6_{\rm fstd} \, T_{\rm f}^{4}\right),
\end{align} 
where $T_{\rm fstd}$ is the standard freeze-out temperature:
 \begin{align}
 \label{eq:Tfstd}
T_{\rm fstd}^6= \frac{\pi^2}{30}\frac{\kappa g_*} {3q^2}\qquad  , \qquad T_{\rm fstd}\sim 0.8 {\rm \,MeV},
\end{align}
corresponding to the $\alpha\rightarrow0$ limit of our model. Solving Eq.~\eqref{eq:modT} for $T_{\rm f}$, we find the freeze-out temperature in qEMSF interaction model:\footnote{Eq.~\eqref{eq:modT} can be expressed as a cubic equation with the redefinition of the variable as $y=T_{\rm f}^2$. In this way, the discriminant of this equation is written as $\Delta=- 27\, T_{\rm fstd}^{12} \left(1+256 \frac{\alpha^3 q^6}{\kappa^3} T_{\rm fstd}^{30} \right)$. Utilizing the constraint in Eq.~\eqref{eq:preliconst2} along with the appropriate constants, it is seen that the second term in brackets is at the order of $\mathcal{O}(10^{-37})$. This indicates that $\Delta<0$ for both $\alpha>0$ and $\alpha<0$, and hence, Eq.~\eqref{eq:modT} has one real root, which corresponds to Eq.~\eqref{Tsol}, and two nonreal complex conjugate roots.}
\begin{equation}
\label{Tsol}
T_{\rm f}=T_{\rm fstd}\left[1+\bar{\alpha} \, T_{\rm fstd}^{6} \, \left(\frac{1}{6}\, k +\frac{1}{3}\, \bar{\alpha} \, T_{\rm fstd}^{12} +\frac{2}{3}\, \bar{\alpha}^2 \, \frac{ T_{\rm fstd}^{24}}{k} \right)^2\right]^{1/6},
\end{equation}
where 
\begin{align}
\label{Tsolterms}
 k=T_{\rm fstd}^{2} \left( 108+8 \,\bar{\alpha}^3 \,T_{\rm fstd}^{30} +12\, \sqrt{81+12\, \bar{\alpha}^3 \,T_{\rm fstd}^{30}}\right)^{1/3},
 \end{align}
with $\bar{\alpha}=\frac{12q^2}{\kappa}\alpha$. It is noteworthy that the freeze-out temperature increases with positive values of $\bar{\alpha}$ and decreases when $\bar{\alpha}$ is negative.

The neutron-to-proton equilibrium ratio at freeze-out can be calculated using:
 \begin{equation}
 \label{eq:neutprot}
\left(\frac{n_{\rm n}}{n_{\rm p}}\right)_{\rm f}\simeq e^{-(m_{\rm n}-m_{\rm p})/{T_{\rm f}}},
 \end{equation}
where $m_{\rm n}-m_{\rm p}=1.293\,{\rm MeV}$ is the mass difference between neutron and proton (determined by both
strong and electromagnetic interactions), and $n_{\rm n}$ and $n_{\rm p}$ represent the neutron and proton number densities, respectively. It is seen that an increase/decrease in $T_{\rm f}$ directly results in a corresponding increase/decrease in the neutron-to-proton ratio when $\alpha$ is positive/negative. The primordial mass
fraction of $^4$He is estimated using~\cite{kolb}:
 \begin{equation}
 \label{Yp}
    Y_{\rm P} \equiv \lambda \, \frac{2 ({n_{\rm n}}/{n_{\rm p}})_{\rm f}}{1+({n_{\rm n}}/{n_{\rm p}})_{\rm f}}\,,
 \end{equation}
where the factor $\lambda$ accounts for the fraction of neutrons that decay into protons (${\rm n}\leftrightarrow {\rm p}+{\rm e}^{-}+\bar{\rm{\nu}}_{{\rm e}}$) during the time interval from $t_{\rm f}$, the freeze-out time of the weak interactions, to $t_{\rm nuc}$, the corresponding freeze-out time of nucleosynthesis. It is expressed as:
 \begin{equation}
 \label{eq:lambda-std}
\lambda_{\rm}=e^{-(t_{\rm nuc}-t_{\rm f})/\tau},
 \end{equation}
 with $\tau=1/\Gamma_{({\rm n} \leftrightarrow {\rm p}+{\rm e}^{-}+\bar{\nu}_{\rm e})}= 887\,{\rm s}$ representing the mean lifetime of a neutron.  Due to the decay of free neutrons after freeze-out, the neutron-to-proton ratio, $({n_{\rm n}}/{n_{\rm p}})_{\rm fstd}\sim 1/5$, slightly decreases to $({n_{\rm n}}/{n_{\rm p}})_{\rm nucstd}\sim 1/7$, resulting in $Y_{\rm P}\sim 0.25$ in standard BBN (SBBN).  

 In the qEMSF interaction model, the estimation of $^4$He abundance is more complicated.\footnote{Unlike models explored in Refs.~\cite{Faulkner:2006ub,Akarsu:2019pvi,Capozziello:2017bxm,Hogas:2021saw,Anagnostopoulos:2022gej,Bhattacharjee,Alvey,Lambiase:2005kb,Barrow:2020kug}, the modification of $^4$He abundance in the qEMSF interaction model can be determined neither by a single parameter $S=H/H_{\rm std}$ that measures the alteration of the expansion rate, nor solely by the shift in the freeze-out temperature denoted as $\delta T_{\rm f}/T_{\rm f}$. This is because the nonconservation of the radiation (in blackbody form) also results in an alteration in the time-temperature relation.} The $^4$He abundance is modified due to two main factors: \textbf{(i)} the neutron-to-proton ratio at freeze-out, $(n_{\rm n}/n_{\rm p})_{\rm f}$, is modified due to the modification in $T_{\rm f}$, as already determined in Eq.\eqref{Tsol}, and \textbf{(ii)} the neutron-to-proton ratio at the onset of nucleosynthesis, $(n_{\rm n}/n_{\rm p})_{\rm nuc}$, is affected both by the modified ratio $(n_{\rm n}/n_{\rm p})_{\rm f}$ and the modification in the time-temperature relation. This relation can be expressed using $a=a_1 t^{1/2}$ in Eq.~\eqref{scale-T} as
\begin{equation}\label{eq:tTrel}
t^2=\frac{T_{\rm std}^4}{T^4} \left(1+ \frac{\bar{\alpha}}{4 q^2 t_{\rm std}^2}\frac{T^4}{T_{\rm std}^4}\right)^{-1}t_{\rm std}^2,
\end{equation}
where we have utilized the standard time-temperature relation:
\begin{equation}
\begin{aligned}
t_{\rm std}^2= \left(\frac{4}{3}\kappa \frac{\pi^2}{30}g_* T_{\rm std}^4\right)^{-1}.
\end{aligned}
\end{equation}
This modification in the time-temperature relation implies a shift in $t_{\rm f}$, the time when $T_{\rm f}$ is reached, and in $t_{\rm nuc}$, the time at which neutrons are captured into deuterium (D).

To perform all the necessary calculations, we utilize an analytic approach for describing freeze-out, as detailed in Refs.~\cite{bbf,muk}. To incorporate the modified temperature relation of the qEMSF interaction model, as outlined in Eq.~\eqref{Tsol}, into our calculations, we substitute it into Eq.~\eqref{eq:tTrel}. This allows us to compute the freeze-out time using the following equation:
\begin{equation}
\begin{aligned}  \label{timef}
\frac{t_{\rm f}}{t_{\rm fstd}}=&(1+6\zeta)^{-1/3} \left[1+\frac{\bar{\alpha}}{4 q^2 t_{\rm fstd}^2}(1+6\zeta)^{2/3}\right]^{-1/2},
\end{aligned}
\end{equation}
where $\zeta$ is a newly defined constant for simplicity, expressed as
\begin{equation}
 \begin{aligned}  \label{zeta}
 \zeta=\frac{\bar{\alpha}}{6}T_{\rm fstd}^6 \left(\frac{1}{6}\, k +\frac{1}{3}\, \bar{\alpha} \, T_{\rm fstd}^{12} +\frac{2}{3}\, \bar{\alpha}^2 \, \frac{ T_{\rm fstd}^{24}}{k} \right)^2, 
  \end{aligned}
 \end{equation}
 thus allowing Eq.~\eqref{Tsol} to be rewritten as ${T_{\rm f}=T_{\rm fstd} (1+6 \zeta)^{1/6}}$. Nucleosynthesis, which occurs subsequent to electron-positron (${\rm e}^{-}{\rm e}^{+}$) annihilation, involves photons and neutrinos as the primary relativistic species. Accordingly, the effective number of relativistic degrees of freedom, $g_*$, is given by $2 + \frac{7}{4}N_\nu (\frac{4}{11})^{4/3}~$\cite{Dodelson03}, reflecting the contributions from photons and neutrinos. For determining the time of nucleosynthesis, $t_{\rm nuc}$, we assume $T_{\rm nuc} = T_{\rm nucstd}$, as the standard nucleosynthesis temperature, $T_{\rm nucstd}$, is predominantly dictated by the D-binding energy. This standard nucleosynthesis temperature, around $0.07\,{\rm MeV}$, marks the onset of deuterium and other light elements' production as per the Standard Model of microphysics~\cite{Torres:1997sn,Lambiase:2005kb,Lambiase:2012fv,Lambiase:2011zz}. However, when we apply this assumption to Eq.~\eqref{eq:tTrel}, it becomes evident that the time of nucleosynthesis, $t_{\rm nuc}$, is still modified due to the qEMSF interaction.  This alteration is quantified as:
\begin{equation}
\begin{aligned}   \label{timenuc}
\frac{t_{\rm nuc}}{t_{\rm nucstd}}=&\left(1+ \frac{\bar{\alpha}}{4 q^2 t_{\rm nucstd}^2}\right)^
{-1/2},
\end{aligned}
\end{equation}
where $t_{\rm nucstd}$ is the standard nucleosynthesis time, and the equation reflects how the qEMSF interaction influences the timing of nucleosynthesis relative to the SBBN.

Finally, since we anticipate only small deviations from the standard values, the modified expression for $\lambda$, as originally defined in Eq.~\eqref{eq:lambda-std}, can be approximated using the standard value $\lambda_{\rm std}=e^{(t_{\rm fstd}-t_{\rm nucstd})/\tau}$. Therefore, applying first-order Taylor series approximation, viz., $(1+x)^n \approx 1+n x$ provided that $x \ll 1$, to Eqs.~\eqref{timef} and~\eqref{timenuc} before substituting into Eq.~\eqref{eq:lambda-std}, this approximation can be expressed as:
 \begin{equation}
 \label{eq:lambda}
\lambda\approx\lambda_{\rm std}\,\exp\bigg\{-\frac{1}{\tau}\left[2\zeta t_{\rm fstd}+\frac{\bar{\alpha}}{8 q^2}\left(\frac{1+2\zeta}{t_{\rm fstd}}-\frac{1}{t_{\rm nucstd}}\right) \right]\bigg\}.
\end{equation}
In the qEMSF interaction model, the shift in the freeze-out temperature, given by $T_{\rm f}=T_{\rm fstd} (1+6 \zeta)^{1/6}$, consequently leads to a deviation in the neutron-to-proton equilibrium ratio at freeze-out, as described by Eq.~\eqref{eq:neutprot} in the standard model. This deviation is represented by  the equation:
 \begin{equation}
 \begin{aligned}   
 \label{eq:neutprotemsf}
\left(\frac{n_{\rm n}}{n_{\rm p}}\right)_{\rm f}=&\left(\frac{n_{\rm n}}{n_{\rm p}}\right)_{\rm fstd}^{1-\zeta},
 \end{aligned}
 \end{equation}
 as indicated above in the factor \textbf{(i)}. Now, we can explicitly formulate Eq.~\eqref{Yp} with these modified values of the fraction $\lambda$ and the neutron-to-proton ratio obtained from Eqs.~\eqref{eq:lambda} and~\eqref{eq:neutprotemsf}, where the modifications traced by $\bar{\alpha}$ are accounted for using $\zeta$ from Eq.~\eqref{zeta}, resulting in:
 \begin{align}
\label{eq:Yp}
Y_{\rm P} \approx Y_{{\rm P}\rm std}&\exp\Bigg[-\frac{2\zeta t_{\rm fstd}+\frac{\bar{\alpha}}{8 q^2}\left(\frac{1+2\zeta}{t_{\rm fstd}}-\frac{1}{t_{\rm nucstd}}\right)}{\tau}\Bigg]\nonumber \\
&\times\frac{\left(\frac{n_{\rm n}}{n_{\rm p}}\right)_{\rm fstd}^{-\zeta}+\left(\frac{n_{\rm n}}{n_{\rm p}}\right)_{\rm fstd}^{1-\zeta}}{1+\left(\frac{n_{\rm n}}{n_{\rm p}}\right)_{\rm fstd}^{1-\zeta}}.
 \end{align}
It is observed that Eq.~\eqref{eq:Yp} differs significantly from the definition~\eqref{Yp} 
in the standard model. This distinction arises because the qEMSF interaction model introduces complex modifications, which impact both the continuity equation and the form of the Friedmann equation in distinct ways, as evident from Eqs.~\eqref{rhorsol1} and~\eqref{eq:rhoprime}, respectively. Furthermore, this dual effect varies across species via the EoS parameter $w$. In contrast, curvature-type alternative theories that still satisfy local energy-momentum conservation, such as Brans–Dicke and $f(R)$ theories, only impact the expansion rate of the Universe via modifications to the Einstein tensor, $G_{\mu\nu}$. These can be interpreted as either a modification of the value of $g_{*}$ or a variation in the gravitational constant $G$, extensively investigated in Refs.~\cite{Kneller,steigman07,Steigman:2012ve}, and parameterized via $S=H/H_{\rm std}$. It is evident that this method is insufficient for investigating modifications arising from EMSF interaction models, or more broadly, from nonminimal matter interaction models. In the next subsection, we will constrain the free parameter $\alpha$ (equivalently $\bar{\alpha}$) using measurements of helium abundances. For a brief review of the measurements of light element abundances, see Ref.~\cite{Workman:2022ynf}. 

We have also summarized the constraints on the $\alpha$ parameter, obtained throughout this paper, including the relevant equations and assumptions, in Table~\ref{table}.

\subsection{Constraints}
\label{sec:constraints}
The primordial helium abundance is best estimated through a cosmological model-independent method, namely, direct measurements provided by observations  of helium and hydrogen recombination lines from metal-poor extragalactic H~II (ionized) regions~\cite{Aver:2015iza,Peimbert}. Regression of the helium mass fraction $Y_{\rm P}$ against the oxygen abundance $n_{\rm O}/n_{\rm H}$ based on the results from Aver \textit{et al.}~\cite{Aver:2015iza} for $15$ objects combined with the results for Leo P in Ref.~\cite{Aver:2020fon} (refer to the BBN section in Ref.~\cite{Workman:2022ynf} for a detailed discussion), suggests the following constraint~\cite{Aver:2020fon};
\begin{equation}
Y_{\rm P}=0.2453 \pm 0.0034\,(68\%\, \rm CL) \,\, \textnormal{Aver \textit{et al.}(2021)}.
\label{eq:Ypalpha}
\end{equation}      
This estimation aligns well with the SBBN value of~\cite{Fields:2019pfx}
\begin{equation}
Y_{\rm P}=0.2469\pm 0.0002\,(68\%\, \rm CL) \,\, \textnormal{Fields \textit{et al.}(2020)},
\label{eq:YpalphaPLK}
\end{equation}
which is based on the Planck-CMB [the Planck power spectra (TT,TE,EE+lowE) and Planck lensing data] measured physical baryon density ($\omega_{\rm b}$), assuming the baseline $\Lambda$CDM model~\cite{Aghanim}. 
It is also consistent with the SBBN-independent -- without using standard BBN theory to relate $(\omega_{\rm b},Y_{\rm P},N_{\nu, \rm {eff}})$ -- Planck-CMB estimation of $Y_{\rm P}=0.239\pm0.013\,(68\%\, \rm CL)$~\cite{Aghanim}, within the baseline $\Lambda$CDM framework, from the combination of Planck power spectra (TT,TE,EE+lowE) and Planck lensing data.

\begin{figure}[t!]
  \includegraphics[width=0.48\textwidth]{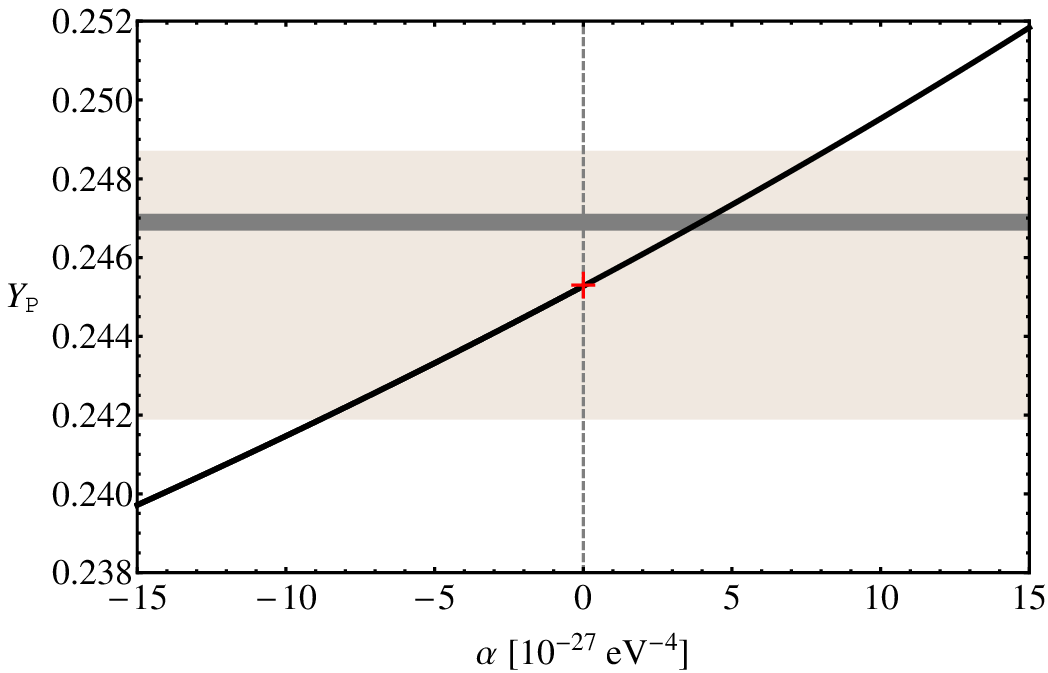}
  \caption{$^4$He mass fraction $Y_{\rm P}$ versus $\alpha$ in the qEMSF interaction model (black line). The red cross represents the SBBN limit corresponding to $\alpha=0$ in the qEMSF interaction model. The wheat and gray bands represent the recent estimations of the primordial helium abundance from H~II regions~\cite{Aver:2020fon} and Planck-$\Lambda$CDM for SBBN~\cite{Fields:2019pfx}, respectively.}
  \label{fig:range}
\end{figure}

Utilizing the latest and widely accepted estimates of the primordial helium abundance from Refs.~\cite{Aver:2020fon,Fields:2019pfx}, as presented in Eqs.~\eqref{eq:Ypalpha} and~\eqref{eq:YpalphaPLK}, the corresponding permissible regions for the free parameter $\alpha$ in the qEMSF interaction model are determined as:
\begin{equation}
\label{He-constraint1}
(-8.81\leq\alpha \leq 8.14)\times 10^{-27} \textnormal{ eV$^{-4}$}
\end{equation}
at $68\%$\, CL for Aver \textit{et al.} (2021)~\cite{Aver:2020fon}, and
\begin{equation} \label{He-constraint2}
(3.48 \leq\alpha \leq 4.43)\,\times 10^{-27} \textnormal{ eV$^{-4}$}
\end{equation}
at $68\%$\, CL for Fields \textit{et al.} (2020)~\cite{Fields:2019pfx}. These constraints are in agreement with, yet more stringent than, our preliminary constraint of ${-10^{-25}\lesssim \alpha \lesssim 10^{-25} \,{\rm eV^{-4}}}$, given in  Eq.~\eqref{eq:preliconst2}. The constraint on $\alpha$ obtained from the measurement reported by Aver~\textit{et al.}, as given in Eq.~\eqref{He-constraint1}, aligns with $\alpha=0$. In contrast, the constraint on $\alpha$ derived from the more stringent $Y_{\rm P}$ estimation reported by Fields \textit{et al.}, as given in Eq.~\eqref{He-constraint2}, suggests a deviation from $\alpha=0$. This outcome lends support to the validity of the qEMSF interaction model, indicating its potential significance in influencing the observed phenomena. Fig.~\ref{fig:range} illustrates the relationship between the helium mass fraction $Y_{\rm P}$ and the parameter $\alpha$ in the qEMSF interaction model. The wheat and gray bands represent recent primordial helium abundance results from H~II Region~\cite{Aver:2020fon} and SBBN+CMB~\cite{Fields:2019pfx}, respectively, considering $N_{\nu,{\rm eff}}=3.046$ with 68\% CL, while the black line demonstrates the increase or decrease in the helium mass fraction corresponding to the increase or decrease in the quadratic radiation correction due to the qEMSF interaction. The ${\alpha=0}$ (i.e., without qEMSF) dotted-gray line corresponds to the standard cosmological model, intersecting the black line at $Y_{\rm P}=0.245$ (red cross).

\begin{table*}[t!]
\caption{The constraints on $\alpha$ parameter obtained throughout the paper.} 
\resizebox{\textwidth}{!}{%
\begin{tabular}{ | l | l | l | }
\hline
\bf{Range of $\alpha$ parameter} $\mathbf{\left[{\rm eV^{-4}}\right]}$ & \bf{Equation} & \bf{Assumption}  %\bf{Parameters }  
\\ \hline  \hline
$- 10^{-22}\lesssim \alpha \lesssim 10^{-22}$ & Friedmann in Eq.~\eqref{eq:rhoprime1}  & minimal deviation from $\rho_{\rm bbn}$  \\   \hline 
$ -1.8\times 10^{-23}\lesssim \alpha \lesssim 6.6\times10^{-12}$ &  radiation domination  at BBN in Eq.~\eqref{rad-dom} & $a_{\rm bbn} \ll 1 $   
  \\ \hline 
$- 10^{-23}\lesssim \alpha \lesssim 10^{-23} $ & $H/H_{\rm std}$ parametrization in Eq.~\eqref{eq:Hubblemod} & $\rho_{\rm r}$ and $\rho_{\rm r}^2$ contribute at the same order  
  \\  \hline
  $ -10^{-25}\lesssim \alpha \lesssim 10^{-25} $ & modified $a-T$ relation in Eq.~\eqref{scale-T} & standard and correction terms are at the same order  
  \\  \hline
 $ (-8.81\leq\alpha \leq 8.14)\times 10^{-27} $ & He-4 mass fraction in Eq.~\eqref{eq:Yp} & $Y_{\rm P}=0.2453 \pm 0.0034\,(68\%\, \rm CL) \,\, \textnormal{Aver \textit{et al.}(2021)}$  
  \\  \hline
   $(3.48 \leq\alpha \leq 4.43)\,\times 10^{-27}$ & He-4 mass fraction in Eq.~\eqref{eq:Yp} & $Y_{\rm P}=0.2469\pm 0.0002\,(68\%\, \rm CL) \,\, \textnormal{Fields \textit{et al.}(2020)}$  
  \\  \hline
   $(-2.17< \alpha < -0.278) \times 10^{-26}$ & He-4 mass fraction in Eq.~\eqref{eq:Yp} & one extra relativistic species, Aver \textit{et al.}(2021)  
  \\  \hline
   $(-7.96< \alpha < -6.90)\times 10^{-26} $ & He-4 mass fraction in Eq.~\eqref{eq:Yp} & one extra relativistic species, Fields \textit{et al.}(2020)  
  \\  \hline
\end{tabular}}
\label{table}
 \end{table*}  

Similar, yet slightly less stringent, constraints on the $\alpha$ parameter of EMSG from BBN, specifically $\alpha \lesssim 6.9 \times 10^{-26}\, \rm{eV^{-4}}$ (where $\alpha$ is assumed to be positive), were also presented in a recent paper on arXiv~\cite{Jang:2024jso}. Unlike our work, which delves into a comprehensive mathematical and physical discussion on the dynamics of BBN in EMSG, this work directly employs the modified Friedmann equation, i.e., $H(z)$, along with an updated and modified version of the Wagoner code and revised reaction rates from the JINA REACLIB Database to derive the constraints. It is worth noting here that these constraints on $\alpha$ are consistent with those that might be obtained by adopting the BBN constraints on brane-world and loop quantum cosmologies, as these models also introduce contributions of quadratic matter terms in the Einstein field equations, similar to the qEMSF interaction model. To be more specific, let us consider the type II Randall–Sundrum brane cosmologies~\cite{Randall:1999vf,Csaki:1999jh,Cline:1999ts,Binetruy:1999hy}, which replace the conventional energy density term $\kappa \rho$ by $\kappa\rho+ \rho^2/(12  M_{5}^6)$ in the Friedmann equation, where the coupling of the quadratic term is related to the brane tension. In Ref.~\cite{Bratt:2002xt}, the five-dimensional Planck mass $M_{5}$ has been constrained using bounds on the abundances of deuterium and $^4$He, and a precise BBN bound has been obtained as $M_5 > 13 \, \mathrm{TeV}$. Eq.~\eqref{eq:rhoprime} establishes a correspondence where $4 \kappa \alpha = 1/(12 M_{5}^6)$, resulting in values smaller than $1.73 \times 10^{-80} \, \mathrm{eV^{-6}}$ when this bound is applied, thus implying that $\alpha \lesssim 2.55 \times 10^{-26} \, \mathrm{eV^{-4}}$ in our model.

In our analysis, we assumed that there are three species of neutrinos (electron, muon and tau) and their antineutrinos, which are left-handed, fully in equilibrium, and each sufficiently low in mass to be highly relativistic at the time of BBN. Any deviation from the standard value of $N_{\nu}=3$ ($N_{\nu, {\rm eff}} = 3.046$ when small corrections for nonequilibrium neutrino heating are included in the thermal evolution), attributed to these three neutrino species, could either indicate the existence of dark radiation (i.e., extra relativistic relics) or arise from modifications induced by the qEMSF interaction, as traced by the parameter $\alpha$. Therefore, the net deviation can be quantified as $\Delta N_{\nu} = \Delta N^{\rm dr}_{\nu}+\Delta N^{\alpha}_{\nu}$ within the qEMSF interaction framework. Various extensions of the SBBN (corresponding to $\Delta N^{\alpha}_{\nu} \rightarrow 0$ and $\alpha \rightarrow 0$ in our model) assume different contributions to $N_{\nu}$ from dark radiation species, such as massless dark gluons ($\Delta N^{\rm dr}_{\nu} \simeq 0.07$)~\cite{Abad}, Goldstone bosons ($\Delta N^{\rm dr}_{\nu} \simeq 0.3$)~\cite{Weinberg02}, and fully thermalized sterile neutrinos ($\Delta N^{\rm dr}_{\nu} = 1.0$)~\cite{Abazajian} as examples. As evident from the modifications in both the modified Friedmann equation~\eqref{eq:rhoprime} and the energy density of radiation~\eqref{scalerho}, the nonminimal interaction with qEMSF can generate a deviation in $N_{\nu}$ even with just three neutrino species, without assuming any dark radiation ($\Delta N^{\rm dr}_{\nu}=0$). Notably, qEMSF interaction with positive $\alpha$ values can mimic the effect of dark radiation.  In other words, an increase in the entropy of usual radiation is akin to an increase in the number of neutrino species. On the other hand, assuming extra relativistic degrees of freedom in the qEMSF model leads to contributions from both $\Delta N^{\rm dr}_{\nu}$ and $\Delta N^{\alpha}_{\nu}$. In general, small deviations from the SBBN scenario are examined through the modified expansion rate of the universe during the BBN epoch, compared to the SBBN value~\cite{Kneller,steigman07,Steigman:2012ve}. This deviation can be quantified as $S=H/H_{\rm std}=\sqrt{\tilde{g}_*/g_*}=\sqrt{1+7/43 \,\Delta N^{\rm dr}_{\nu}}$,  where $\tilde{g}_{*}$ represents the modified effective number of relativistic species determining the radiation energy density. This modification impacts the $^4$He fraction, following the relation  $Y_{\rm P}=0.2381\pm0.0006+0.0016[\eta_{10}+100 (S-1)]$. Here, $\eta_{10}=273.9\, \Omega_{\rm b0}h^2\approx6.1$, using the mean value of  $\Omega_{\rm b0}h^2=0.02244$ from observational analysis within the $\Lambda$CDM model. For instance, extending the SBBN scenario to include fully thermalized sterile neutrinos with $\Delta N^{\rm dr}_{\nu}= 1.0$~\cite{Abazajian} results in $Y_{\rm P}=0.2604 \pm 0.0006$, conflicting with the observational values in Eqs.~\eqref{eq:Ypalpha} and~\eqref{eq:YpalphaPLK}. However, within the qEMSF interaction framework, such deviations are analyzed using the method presented in Subsection~\ref{sec:prinuc}, with $S$ serving to trace expansion rate deviations. Consequently, it is possible to simultaneously consider the existence of an additional degree of freedom (e.g., sterile neutrino contributing $\Delta N^{\rm dr}_{\nu} = 1.0$~\cite{Abazajian}) and bring the primordial $^4$He mass fraction into alignment with observed values through the contributions from $\alpha$, viz., $\Delta N^{\alpha}_{\nu}$, within the following ranges:
\begin{equation}
(-2.17< \alpha < -0.278) \times 10^{-26} \textnormal{ eV$^{-4}$}
\end{equation}
at $68\%$\, CL for Aver \textit{et al.} (2021)~\cite{Aver:2020fon}, and
\begin{equation}
(-7.96< \alpha < -6.90)\times 10^{-26} \textnormal{ eV$^{-4}$}
\end{equation}
at $68\%$\, CL for Fields \textit{et al.} (2020)~\cite{Fields:2019pfx}. This outcome suggests that the qEMSF interaction model can accommodate an additional neutrino species or a deviation from the Standard Model of particle physics while maintaining the $^4$He mass fraction in line with observational estimates.

\section{Conclusion}
\label{conclusion}
We have explored an extension of the standard cosmological model within the framework of the quadratic energy-momentum squared gravity (qEMSG)~\cite{Roshan:2016mbt,Akarsu:2017ohj,Board:2017ign}, considering it from nonminimal interaction perspective presented in Ref.~\cite{Akarsu:2023lre} regarding matter-type modified theories of gravity. In this model, the usual material field, $T_{\mu\nu}$, described by the matter Lagrangian density, $\mathcal{L}_{\rm m}$, interacts nonminimally with its accompanying quadratic energy-momentum squared field (qEMSF), $T_{\mu\nu}^{\rm qEMSF}$, described by the function $f(\bf T^{2})=\alpha \bf{T^{2}}$ where $\mathbf{T^2}=T_{\mu\nu}T^{\mu\nu}$ and $\alpha$ is the constant free parameter that quantifies the amount of qEMSF with respect to $T_{\mu\nu}$. The strongest astrophysical constraint on $\alpha$, obtained from neutron stars, is $|\alpha|\lesssim  10^{-35} \,{\rm eV^{-4}}$ as indicated in Ref.~\cite{Akarsu:2018zxl}. This significant constraint arises due to the enhanced impact of quadratic material corrections from qEMSF at higher energy densities. Neutron stars, being among the densest astrophysical objects, thus provide the most stringent astrophysical constraints. Similarly, strong cosmological constraints on $\alpha$ can be derived by examining phenomena in the early Universe, where the density was significantly greater than it is today. Therefore, in this work, we have stringently constrained the $\alpha$ parameter from cosmology, focusing on the early Universe, particularly the BBN epoch.

In our model's cosmological context, pressureless matter adheres to local energy-momentum conservation, whereas the continuity equation for radiation incorporates an additional term. As demonstrated first in Ref.~\cite{Akarsu:2018zxl}, one of the two solutions derived from the modified Friedmann equations fails to converge to the standard cosmological model in the limit of $\alpha \rightarrow 0$, which is expected to nullify the qEMSF corrections. Selecting the solution that aligns with the standard cosmological model in the $\alpha \rightarrow 0$ limit, we employed the thermodynamics of open systems to analyze the radiation component in our model. We demonstrated that the energy-momentum/entropy transfer exhibits a bidirectional nature dependent on the sign of $\alpha$: transfer occurs from the qEMSF counterpart to the usual radiation component for $\alpha>0$, and conversely for $\alpha<0$. The qEMSF interaction model, characterized by the nonconservation of energy-momentum, alters the time/scale factor dependence of energy density in the radiation-dominated era, diverging from the standard cosmological model. Despite this, the scale factor-time relation remains unaffected due to the deviation in the expansion rate, expressed as $H(\rho_{\rm r})/H_{\rm std}(\rho_{\rm r})=\sqrt{1+4\alpha \rho_{\rm r}}$. Moreover, the modified radiation energy density evolution in qEMSF interaction still adheres to the standard blackbody radiation form, $\rho_{\rm r}\propto T^4$,  leading to a change in the standard time/scale factor-temperature relation. By applying these unique features of our model via a semi-analytical approach, we uncover complex effects on the abundances of the primordial light elements formed during the BBN, diverging from the predictions of the SBBN. Specifically, positive values of $\alpha$ speeds up the freeze-out process of neutron--proton interconversion, consequently increasing the $^4$He mass fraction, while negative values of $\alpha$ have the inverse effect. We have derived the most stringent cosmological constraints on the parameter $\alpha$ to data, which are $(-8.81\leq\alpha \leq 8.14)\times 10^{-27} \rm{eV}^{-4}$ at $68\%$\, CL, using the most recent primordial helium abundance measurements from Aver~\textit{et al.}~\cite{Aver:2020fon}, and $(3.48 \leq\alpha \leq 4.43)\,\times 10^{-27} \rm{eV}^{-4}$ at $68\%$\, CL, using the SBBN estimation of helium abundance, utilizing the physical baryon density estimation from the Planck-CMB observations assuming the baseline $\Lambda$CDM model, given in Fields \textit{et al.}~\cite{Fields:2019pfx}. Importantly, the range derived from Aver~\textit{et al.}'s measurements aligns with $\alpha=0$, suggesting no evidence for favoring the qEMSF interaction model over the standard model. Conversely, the range derived from Fields \textit{et al.}'s estimation, which does not encompass $\alpha=0$, lends support to the validity of the qEMSF interaction model. This indicates its potential significance in influencing the observed phenomena. It has been observed that positive values of $\alpha$ in the qEMSF interaction model mimic the effects of dark radiation, effectively increasing the total number of massless degrees of freedom. Consequently, the qEMSF interaction model with negative $\alpha$ values could be used to counterbalance extra degrees of freedom that might arise due to the potential presence of, for instance, a sterile neutrino. In light of this, we have investigated the constraints on $\alpha$ assuming one extra relativistic degree of freedom exists alongside qEMSF. We determined that $(-2.17< \alpha < -0.278) \times 10^{-26} \, \rm{eV}^4$ at $68\%$\, CL is the constraint for the $^4$He mass fraction measurement in Aver \textit{et al.}~\cite{Aver:2020fon}, and $(-7.96< \alpha < -6.90)\times 10^{-26}\, \rm{eV}^4$ at $68\%$\, CL for the value recommended by Fields \textit{et al.}~\cite{Fields:2019pfx}. These results indicate that the qEMSF interaction model has the potential to reconcile deviations from standard cosmology and the Standard Model of elementary particles, bringing them into alignment with primordial helium abundance measurements.

In this work, anticipating the most stringent cosmological limits from BBN, we utilized the primordial $^4$He abundance to constrain the parameter $\alpha$ in the qEMSF interaction model. The constraints obtained can be further refined with the anticipated improvements from the CMB-S4 measurements---the next generation ground-based CMB measurements~\cite{CMB-S4:2016ple,CMB-S4:2022ght}. As temperatures fall below $0.1$ MeV, another crucial reaction to consider is the production of deuterium (${\rm n}+{\rm p}\leftrightarrow {\rm D}+ {\rm \gamma}$). Deuterium's abundance decreases monotonically in the post-BBN Universe due to its complete destruction during stellar evolution, rendering BBN the sole significant source of this element (see Ref.~\cite{prodanovic} and references therein). This unique aspect of deuterium provides an opportunity for future research. By employing the primordial abundances of deuterium, further refinement of the constraint on $\alpha$ can be achieved using the semi-analytical method outlined. The most recent astrophysical estimates of the primordial deuterium abundance, derived from the best seven measurements in metal-poor damped  Lyman-$\alpha$ systems, suggest a mass fraction of $y_{\rm DP} \equiv 10^5\, n_{\rm D}/n_{\rm H}=2.527 \pm 0.030$, where $n_{\rm D}$ and $n_{\rm H}$ are the number densities of deuterium and hydrogen, respectively~\cite{Cooke:2017cwo}. This highlights deuterium as the most constraining primordial element among all, with its observational measurement reaching 1\% accuracy~\cite{Workman:2022ynf,Mossa:2020gjc,Pitrou:2020etk,Pisanti:2020efz,Yeh:2020mgl}. Therefore, to advance our current analysis, the primordial abundances of deuterium can be utilized to improve the constraint on $\alpha$ by following the analytical method presented for the qEMSF interaction model.\\

\newpage

\begin{acknowledgments}
We extend our heartfelt gratitude to Professor John David Barrow, whose invaluable contributions and guidance at the inception of this project continue to inspire us following his passing. \"{O}.A. acknowledges the support by the Turkish Academy of Sciences in scheme of the Outstanding Young Scientist Award (T\"{U}BA-GEB\.{I}P). M.B.L. is supported by the Basque Foundation of Science Ikerbasque and has
been financed by the Spanish project PID2020-114035GB-100 (MINECO/AEI/FEDER, UE). M.B.L. also
would like to acknowledge the financial support from the Basque government Grant No. IT1628-22
(Spain). N.K. thanks Do\u gu\c s University for the financial support provided by the Scientific Research (BAP) project number 2021-22-D1-B01. \"O.A., N.K., and N.M.U. are supported in part by T\"UB\.ITAK grant 122F124.  This article is based upon work from COST Action CA21136 Addressing observational tensions in cosmology with systematics and fundamental physics (CosmoVerse) supported by COST (European Cooperation in Science and Technology).

\end{acknowledgments}


\begin{thebibliography}{0}%
\makeatletter
\providecommand \@ifxundefined [1]{%
 \@ifx{#1\undefined}
}%
\providecommand \@ifnum [1]{%
 \ifnum #1\expandafter \@firstoftwo
 \else \expandafter \@secondoftwo
 \fi
}%
\providecommand \@ifx [1]{%
 \ifx #1\expandafter \@firstoftwo
 \else \expandafter \@secondoftwo
 \fi
}%
\providecommand \natexlab [1]{#1}%
\providecommand \enquote  [1]{``#1''}%
\providecommand \bibnamefont  [1]{#1}%
\providecommand \bibfnamefont [1]{#1}%
\providecommand \citenamefont [1]{#1}%
\providecommand \href@noop [0]{\@secondoftwo}%
\providecommand \href [0]{\begingroup \@sanitize@url \@href}%
\providecommand \@href[1]{\@@startlink{#1}\@@href}%
\providecommand \@@href[1]{\endgroup#1\@@endlink}%
\providecommand \@sanitize@url [0]{\catcode `\\12\catcode `\$12\catcode `\&12\catcode `\#12\catcode `\^12\catcode `\_12\catcode `\%12\relax}%
\providecommand \@@startlink[1]{}%
\providecommand \@@endlink[0]{}%
\providecommand \url  [0]{\begingroup\@sanitize@url \@url }%
\providecommand \@url [1]{\endgroup\@href {#1}{\urlprefix }}%
\providecommand \urlprefix  [0]{URL }%
\providecommand \Eprint [0]{\href }%
\providecommand \doibase [0]{https://doi.org/}%
\providecommand \selectlanguage [0]{\@gobble}%
\providecommand \bibinfo  [0]{\@secondoftwo}%
\providecommand \bibfield  [0]{\@secondoftwo}%
\providecommand \translation [1]{[#1]}%
\providecommand \BibitemOpen [0]{}%
\providecommand \bibitemStop [0]{}%
\providecommand \bibitemNoStop [0]{.\EOS\space}%
\providecommand \EOS [0]{\spacefactor3000\relax}%
\providecommand \BibitemShut  [1]{\csname bibitem#1\endcsname}%
\let\auto@bib@innerbib\@empty
%</preamble>
\end{thebibliography}%


\begin{thebibliography}{99} 




\bibitem{Katirci:2014sti}
N.~Kat{\i}rc{\i} and M.~Kavuk, $f(R,T_{\mu\nu}T^{\mu\nu})$ gravity and Cardassian-like expansion as one of its consequences, \href{https://doi.org/10.1140/epjp/i2014-14163-6}{Eur. Phys. J. Plus \textbf{129}, 163 (2014)}. \href{https://arxiv.org/abs/1302.4300}{1302.4300}

\bibitem{Roshan:2016mbt}
M.~Roshan and F.~Shojai, Energy-momentum squared gravity, \href{https://doi.org/10.1103/PhysRevD.94.044002}{Phys. Rev. D \textbf{94}, 044002 (2016)}. \href{https://arxiv.org/abs/1607.06049}{1607.06049}
    
\bibitem{Akarsu:2017ohj}
\" O.~Akarsu, N.~Kat{\i}rc{\i} and S.~Kumar, Cosmic acceleration in a dust only universe via energy-momentum powered gravity, \href{https://doi.org/10.1103/PhysRevD.97.024011}{Phys. Rev. D \textbf{97}, 024011 (2018)}. \href{https://arxiv.org/abs/1709.02367}{1709.02367}

\bibitem{Board:2017ign}
C.V.R.~Board and J.D.~Barrow, Cosmological models in energy-momentum-squared gravity, \href{https://doi.org/10.1103/PhysRevD.96.123517}{Phys. Rev. D \textbf{96}, 123517 (2017)}. \href{https://arxiv.org/abs/1709.09501}{1709.09501}

\bibitem{Akarsu:2023lre}
\"O.~Akarsu, M.~Bouhmadi-L\'opez, N.~Kat\i{}rc\i{}, E.~Nazari, M.~Roshan and N.M.~Uzun, Equivalence of matter-type modified gravity theories to general relativity with nonminimal matter interaction, Phys. Rev. D (2024) in press,
\href{https://arxiv.org/abs/2306.11717}{2306.11717}

\bibitem{Harko:2010mv}
T.~Harko and F.S.N.~Lobo, $f(R,L_{m})$ gravity, \href{https://doi.org/10.1140/epjc/s10052-010-1467-3}{Eur. Phys. J. C \textbf{70}, 373 (2010)}. \href{https://arxiv.org/abs/1008.4193}{1008.4193}

\bibitem{Harko:2011kv}
T.~Harko, F.S.N.~Lobo, S.~Nojiri and S.D.~Odintsov, $f(R,T)$ gravity, \href{https://doi.org/10.1103/PhysRevD.84.024020}{Phys. Rev. D \textbf{84},  024020 (2011).} \href{https://arxiv.org/abs/1104.2669}{1104.2669}



\bibitem{Kolonia:2022jje}
E.A.~Kolonia and C.J.A.P.~Martins, Observational constraints on nonlinear matter extensions of general relativity: Separable trace power models, 
\href{https://doi.org/10.1016/j.dark.2022.101021}{Phys. Dark Univ. \textbf{36}  101021 (2022)}.
\href{https://arxiv.org/abs/2204.08016}{2204.08016}

\bibitem{Randall:1999vf}
L.~Randall and R.~Sundrum, An Alternative to compactification, \href{https://doi.org/10.1103/PhysRevLett.83.4690}{Phys. Rev. Lett. \textbf{83} (1999)}.
\href{https://arxiv.org/abs/hep-th/9906064}{hep-th/9906064}

\bibitem{Csaki:1999jh}
C.~Csaki, M.~Graesser, C.F.~Kolda and J.~Terning, Cosmology of one extra dimension with localized gravity,
\href{https://doi.org/10.1016/S0370-2693(99)00896-5}{Phys. Lett. B \textbf{462} (1999)}.
\href{https://arxiv.org/abs/hep-ph/9906513}{hep-ph/9906513}

\bibitem{Cline:1999ts}
J.M.~Cline, C.~Grojean and G.~Servant, Cosmological expansion in the presence of extra dimensions,
\href{https://doi.org/10.1103/PhysRevLett.83.4245}{Phys. Rev. Lett. \textbf{83} 4245 (1999)}. 
\href{https://arxiv.org/abs/hep-ph/9906523}{hep-ph/9906523}


\bibitem{Binetruy:1999hy}
P.~Binetruy, C.~Deffayet, U.~Ellwanger and D.~Langlois, Brane cosmological evolution in a bulk with cosmological constant, 
\href{https://doi.org/10.1016/S0370-2693(00)00204-5}{Phys. Lett. B \textbf{477} (2000)}.
\href{https://arxiv.org/abs/hep-th/9910219}{hep-th/9910219}



\bibitem{Brax:2003fv} 
P.~Brax and C.~van de Bruck, Cosmology and brane worlds: A Review, \href{https://doi.org/10.1088/0264-9381/20/9/202}{Classical Quantum Gravity {\bf 20}, R201 (2003)}. \href{https://arxiv.org/abs/hep-th/0303095}{hep-th/0303095}

\bibitem{Ashtekar:2011ni} 
A.~Ashtekar and P.~Singh, Loop Quantum Cosmology: A Status Report, \href{https://doi.org/10.1088/0264-9381/28/21/213001}{Classical Quantum Gravity \textbf{28}, 213001 (2011)}. \href{https://arxiv.org/abs/1108.0893}{1108.0893}



\bibitem{Barrow:2020tzx}
J.D.~Barrow, The Area of a Rough Black Hole,
\href{https://doi.org/10.1016/j.physletb.2020.135643}{Phys. Lett. B \textbf{808}, 135643 (2020)}.
\href{https://arxiv.org/abs/2004.09444}{2004.09444}
 
\bibitem{Sheykhi:2021fwh}
A.~Sheykhi, Barrow Entropy Corrections to Friedmann Equations, 
\href{https://doi.org/10.1103/PhysRevD.103.123503}{Phys. Rev. D \textbf{103}, 123503 (2021)}. \href{https://arxiv.org/abs/2102.06550}{2102.06550}

\bibitem{Chen:2019dip}
C.Y.~Chen and P.~Chen, Eikonal black hole ringings in generalized energy-momentum squared gravity, \href{https://doi.org/10.1103/PhysRevD.101.064021}{Phys. Rev. D \textbf{101}, 064021 (2020)}. \href{https://arxiv.org/abs/1910.12262}{1910.12262}

\bibitem{Chen:2021cts}
C.Y.~Chen, M.~Bouhmadi-L\'opez and P.~Chen, Lessons from black hole quasinormal modes in modified gravity, \href{https://doi.org/10.1140/epjp/s13360-021-01227-z}{Eur. Phys. J. Plus \textbf{136} 253 (2021)}.  \href{https://arxiv.org/abs/2103.01249}{2103.01249}

\bibitem{Akarsu:2018zxl}
\" O.~Akarsu, J.D.~Barrow, S.~\c C{\i}k{\i}nto\u glu, K.Y.~Ek\c si and N.~Kat{\i}rc{\i}, Constraint on energy-momentum squared gravity from neutron stars and its cosmological implications, \href{https://doi.org/10.1103/PhysRevD.97.124017}{Phys. Rev. D \textbf{97}, 124017 (2018)}. \href{https://arxiv.org/abs/1802.02093}{1802.02093}


\bibitem{Nazari:2022xhv}
E.~Nazari, M.~Roshan and I.~De Martino,
Constraining energy-momentum-squared gravity by binary pulsar observations,
\href{https://doi.org/10.1103/PhysRevD.105.044014}{Phys. Rev. D \textbf{105}, 044014 (2022)}.
\href{https://arxiv.org/abs/2201.08578}{2201.08578}

\bibitem{Nari:2018aqs} 
N.~Nari and M.~Roshan, Compact stars in energy-momentum squared gravity, \href{https://doi.org/10.1103/PhysRevD.98.024031}{Phys. Rev. D {\bf 98}, 024031 (2018)}. \href{https://arxiv.org/abs/1802.02399}{1802.02399}


\bibitem{HosseiniMansoori:2023zop}
S.A.~Hosseini Mansoori, F.~Felegary, M.~Roshan, \" O.~Akarsu and M.~Sami, $T^2$- inflation: Sourced by energy-momentum squared gravity, \href{https://doi.org/10.1016/j.dark.2023.101360}{Phys. Dark Univ. \textbf{42}, 101360 (2023)}.
\href{https://arxiv.org/abs/2306.09181}{2306.09181}


\bibitem{HosseiniMansoori:2023mqh}
S.A.~Hosseini Mansoori, F.~Felegray, A.~Talebian and M.~Sami,
PBHs and GWs from \ensuremath{\mathbb{T}}$^{2}$-inflation and NANOGrav 15-year data,
\href{https://doi.org/10.1088/1475-7516/2023/08/067}{J. Cosmol. Astropart. Phys. \textbf{08}, 067 (2023)}.
\href{https://arxiv.org/abs/2307.06757}{2307.06757} 

\bibitem{bbf} 
J.~Bernstein, L.S.~Brown and G.~Feinberg, Cosmological Helium Production Simplified, 
\href{https://doi.org/10.1103/RevModPhys.61.25}{Rev. Mod. Phys. {\bf 61}, 25 (1989)}.

\bibitem{kolb}
E.W.~Kolb and M.S.~Turner, {\it The Early Universe }, (Addison Wesley
Publishing Company, Redwood City, USA, 1989).

\bibitem{Olive:1999ij}
K.A.~Olive, G.~Steigman and T.P.~Walker, Primordial nucleosynthesis: Theory and observations, 
\href{https://doi.org/10.1016/S0370-1573(00)00031-4}{Phys. Rep. \textbf{333}, 389 (2000)}.
\href{https://arxiv.org/abs/astro-ph/9905320}{astro-ph/9905320}

\bibitem{Cyburt:2015mya}
R.H.~Cyburt, B.D.~Fields, K.A.~Olive and T.H.~Yeh, Big Bang Nucleosynthesis: 2015,
\href{https://doi.org/10.1103/RevModPhys.88.015004}{Rev. Mod. Phys. \textbf{88},  015004 (2016)}. \href{https://arxiv.org/abs/1505.01076}{1505.01076}

\bibitem{muk} 
V.F.~Mukhanov, Nucleosynthesis without a computer, \href{https://doi.org/10.1023/B:IJTP.0000048169.69609.77}{Internat. J. Theoret. Phys.  {\bf 43}, 669 (2004)}.
\href{https://arxiv.org/abs/astro-ph/0303073}{astro-ph/0303073}

\bibitem{Baumannbook}
D.~Baumann, \textit{Cosmology}, (Cambridge University Press, Cambridge, England, 2022).

\bibitem{Faulkner:2006ub}
T.~Faulkner, M.~Tegmark, E.F.~Bunn and Y.~Mao, Constraining f(R) Gravity as a Scalar Tensor Theory, \href{https://doi.org/10.1103/PhysRevD.76.063505}{Phys. Rev. D \textbf{76},  063505 (2007)}. 
\href{https://arxiv.org/abs/astro-ph/0612569}{astro-ph/0612569}

\bibitem{Akarsu:2019pvi}
\" O.~Akarsu, N.~Kat\i{}rc\i{}, N.~\"Ozdemir and J.A.~V\'azquez, Anisotropic massive Brans-Dicke gravity extension of the standard $\Lambda$CDM model, 
\href{https://doi.org/10.1140/epjc/s10052-019-7580-z}{Eur. Phys. J. C \textbf{80}, 32 (2020)}. 
\href{https://arxiv.org/abs/1903.06679}{1903.06679}

\bibitem{Capozziello:2017bxm}
S.~Capozziello, G.~Lambiase and E.N.~Saridakis, Constraining $f(T)$ teleparallel gravity by Big Bang Nucleosynthesis, 
\href{https://doi.org/10.1140/epjc/s10052-017-5143-8}{Eur. Phys. J. C \textbf{77}, 576 (2017)}.
\href{https://arxiv.org/abs/1702.07952}{1702.07952}

\bibitem{Hogas:2021saw}
M.~H\"og\r{a}s and E.~M\"ortsell, Constraints on bimetric gravity from Big Bang nucleosynthesis, 
\href{https://doi.org/10.1088/1475-7516/2021/11/001}{J. Cosmol. Astropart. P. \textbf{11}, 001 (2021)}.
\href{https://arxiv.org/abs/2106.09030}{2106.09030}

\bibitem{Anagnostopoulos:2022gej}
F.K.~Anagnostopoulos, V.~Gakis, E.N.~Saridakis and S.~Basilakos,
New models and big bang nucleosynthesis constraints in f(Q) gravity,
\href{https://doi.org/10.1140/epjc/s10052-023-11190-x}{Eur. Phys. J. C \textbf{83}, 58 (2023)}.
\href{https://arxiv.org/abs/2205.11445}{2205.11445}

\bibitem{Bhattacharjee} 
S.~Bhattacharjee and P.K.~Sahoo, Big bang nucleosynthesis and entropy evolution in $f(R,T)$ gravity, 
\href{https://doi.org/10.1140/epjp/s13360-020-00361-4}{Eur. Phys. J. Plus \textbf{135}, 350 (2020)}.
\href{https://arxiv.org/abs/2004.04684}{2004.04684}

\bibitem{Alvey} 
J.~Alvey, N.~Sabti, M.~Escudero \textit{et al.}, 
Improved BBN constraints on the variation of the gravitational constant, 
\href{https://doi.org/10.1140/epjc/s10052-020-7727-y}{Eur. Phys. J. C \textbf{80}, 148 (2020)}. 
\href{https://arxiv.org/abs/1910.10730}{1910.10730}

\bibitem{Lambiase:2005kb}
G.~Lambiase, Lorentz invariance breakdown and constraints from big-bang nucleosynthesis,
\href{https://doi.org/10.1103/PhysRevD.72.087702}{Phys. Rev. D \textbf{72}, 087702 (2005)}.
\href{https://arxiv.org/abs/astro-ph/0510386}{astro-ph/0510386}

\bibitem{Barrow:2020kug}
J.D.~Barrow, S.~Basilakos and E.N.~Saridakis, Big Bang Nucleosynthesis constraints on Barrow entropy, \href{https://doi.org/10.1016/j.physletb.2021.136134}{Phys. Lett. B \textbf{815},  136134 (2021)}. 
\href{https://arxiv.org/abs/2010.00986}{2010.00986}

\bibitem{Jang:2024jso}
D.~Jang, M.R.~Gangopadhyay, M.K.~Cheoun, T.~Kajino and M.~Sami,
Big Bang Nucleosynthesis constraints on the Energy-Momentum Squared Gravity: The $\mathbb{T}^{2}$ model,
\href{https://arxiv.org/abs/2402.01210}{2402.01210}


\bibitem{Lovelock:1971yv} D.~Lovelock, The Einstein tensor and its generalizations, 
\href{https://doi.org/10.1063/1.1665613}{J. Math. Phys. \textbf{12}, 498 (1971)}.

\bibitem{Lovelock:1972vz} D.~Lovelock, The four-dimensionality of space and the Einstein tensor, 
\href{https://doi.org/10.1063/1.1666069}{J. Math. Phys. \textbf{13}, 874 (1972)}.

\bibitem{Brown:1993}
J.D.~Brown, Action functionals for relativistic perfect fluids, \href{https://doi.org/10.1088/0264-9381/10/8/017}{Class. Quantum Gravity \textbf{10}, 1579 (1993).}
\href{https://arxiv.org/abs/gr-qc/9304026}{gr-qc/9304026}

\bibitem{Taub:1954}
A.H.~Taub, General Relativistic Variational Principle for Perfect Fluids, \href{https://doi.org/10.1103/PhysRev.94.1468}{Phys. Rev. \textbf{94}, 1468 (1954).}

\bibitem{Schutz1970}
B.F.~Schutz,
Perfect Fluids in General Relativity: Velocity Potentials and a Variational Principle,
\href{https://doi.org/10.1103/PhysRevD.2.2762}{Phys. Rev. D \textbf{2}, 2762 (1970)}.

\bibitem{Haghani:2013oma}
Z.~Haghani, T.~Harko, F.S.N.~Lobo, H.R.~Sepangi and S.~Shahidi,
Further matters in space-time geometry: $f(R,T,R_{\mu\nu}T^{\mu\nu})$ gravity,
\href{https://doi.org/10.1103/PhysRevD.88.044023}{Phys. Rev. D \textbf{88}, 044023 (2013)}.
\href{https://arxiv.org/abs/1304.5957}{1304.5957}

\bibitem{Odintsov:2013iba}
S.D.~Odintsov and D.~S\'aez-G\'omez,
$f(R, T, R_{\mu\nu} T^{\mu\nu})$ gravity phenomenology and $\Lambda$CDM universe,
\href{https://doi.org/10.1016/j.physletb.2013.07.026}{Phys. Lett. B \textbf{725}, 437 (2013)}.
\href{https://arxiv.org/abs/1304.5411}{1304.5411}

\bibitem{Asimakis:2022jel}
P.~Asimakis, S.~Basilakos, A.~Lymperis, M.~Petronikolou and E.N.~Saridakis,
Modified gravity and cosmology with nonminimal (derivative) coupling between matter and the Einstein tensor, \href{https://doi.org/10.1103/PhysRevD.107.104006}{Phys. Rev. D \textbf{107}, 104006 (2023)}.
\href{https://arxiv.org/abs/2212.03821}{2212.03821}

\bibitem{Akarsu:2018aro}
\"{O}.~Akarsu, N.~Kat{\i}rc{\i}, S.~Kumar, R.C.~Nunes and M.~Sami, Cosmological implications of scale-independent energy-momentum squared gravity: Pseudo nonminimal interactions in dark matter and relativistic relics, \href{https://doi.org/10.1103/PhysRevD.98.063522}{Phys. Rev. D {\bf 98}, 063522 (2018)}. \href{https://arxiv.org/abs/1807.01588}{1807.01588}

\bibitem{Akarsu:2020vii}
\"O.~Akarsu, J.D.~Barrow and N.M.~Uzun,
Screening anisotropy via energy-momentum squared gravity: $\Lambda$CDM model with hidden anisotropy,
\href{https://doi.org/10.1103/PhysRevD.102.124059}{Phys. Rev. D \textbf{102}, 124059 (2020)}.
\href{https://arxiv.org/abs/2009.06517}{2009.06517}

\bibitem{Planck15Cosmo} 
 P.A.R.~Ade \textit{et al.} (Planck Collaboration), Planck 2015 results. XIII. Cosmological parameters, \href{https://doi.org/10.1051/0004-6361/201525830}{Astron. Astrophys. \textbf{594}, A13 (2016)}. \href{https://arxiv.org/abs/1502.01589}{1502.01589}

\bibitem{Dodelson03} 
S.~Dodelson and F.~Schmidt, \textit{Modern Cosmology}, (Academic Press, Elsevier Science, New York, 2021).

\bibitem{Shapiro83} 
S.L.~Shapiro and S.A.~Teukolsky, \textit{Black holes, white dwarfs, and neutron stars: The physics of compact objects}, (Wiley, 1983).

\bibitem{Chavanis:2014lra} 
  P.H.~Chavanis, Cosmology with a stiff matter era, \href{https://doi.org/10.1103/PhysRevD.92.103004}{Phys. Rev. D {\bf 92}, 103004 (2015).} \href{https://arxiv.org/abs/1412.0743}{1412.0743}

\bibitem{Barrow78} 
J.D.~Barrow, Quiescent cosmology, 
\href{https://doi.org/10.1038/272211a0}{Nature (London) \textbf{272}, 211 (1978)}.

\bibitem{Akarsu:2019pwn}
\"O.~Akarsu, S.~Kumar, S.~Sharma and L.~Tedesco, Constraints on a Bianchi type I spacetime extension of the standard $\Lambda$CDM model, 
\href{https://doi.org/10.1103/PhysRevD.100.023532}{Phys. Rev. D \textbf{100}, 023532 (2019)}. 
\href{https://arxiv.org/abs/1905.06949}{1905.06949}

 
\bibitem{Akarsu:2021max}
\" O.~Akarsu, E.~Di Valentino, S.~Kumar, M.~Ozyigit and S.~Sharma, Testing spatial curvature and anisotropic expansion on top of the \ensuremath{\Lambda}CDM model, \href{https://doi.org/10.1016/j.dark.2022.101162}{Phys. Dark Univ. \textbf{39} 101162 (2023).}
\href{https://arxiv.org/abs/2112.07807}{2112.07807}

\bibitem{Poulin:2018cxd}
V.~Poulin, T.L.~Smith, T.~Karwal and M.~Kamionkowski, Early Dark Energy Can Resolve The Hubble Tension, \href{https://doi.org/10.1103/PhysRevLett.122.221301}{Phys. Rev. Lett. \textbf{122} 221301 (2019)}.\href{https://arxiv.org/abs/1811.04083}{1811.04083}

\bibitem{DiValentino:2021izs}
E.~Di Valentino, O.~Mena, S.~Pan, L.~Visinelli, W.~Yang, A.~Melchiorri, D.F.~Mota, A.G.~Riess and J.~Silk, In the realm of the Hubble tension \textemdash{}a review of solutions, \href{https://doi.org/10.1088/1361-6382/ac086d}{Classical Quantum Gravity \textbf{38} 153001 (2021)}. \href{https://arxiv.org/abs/2103.01183}{2103.01183}


\bibitem{Barrow:1997sy} 
J.D.~Barrow, Cosmological limits on slightly skew stresses, \href{https://doi.org/10.1103/PhysRevD.55.7451}{Phys. Rev. D {\bf 55}, 7451 (1997).} \href{https://arxiv.org/abs/gr-qc/9701038}{gr-qc/9701038}

\bibitem{Akarsu:2020pka}
O.~Akarsu, N.~Katirci, A.A.~Sen and J.A.~Vazquez,
Scalar field emulator via anisotropically deformed vacuum energy: Application to dark energy, \href{https://arxiv.org/abs/2004.14863}{2004.14863}

 \bibitem{Aghanim} 
N.~Aghanim \textit{et al.} (Planck collaboration),
Planck 2018 results. VI. Cosmological parameters,
\href{https://doi.org/10.1051/0004-6361/201833910}{Astron. Astrophys. \textbf{641}, A6 (2020)},
[erratum: \href{https://doi.org/10.1051/0004-6361/201833910e}{Astron. Astrophys. \textbf{652}, C4 (2021)}].
\href{https://arxiv.org/abs/1807.06209}{1807.06209}


\bibitem{Prigogine:1989zz}
I.~Prigogine, J.~Geheniau, E.~Gunzig and  P.~Nardone, Thermodynamics and cosmology,
\href{https://doi.org/10.1007/BF00758981}{Gen. Relativity Gravitation \textbf{21}, 767 (1989)}.

\bibitem{Fixsen:2009ug}
D.J.~Fixsen,
The Temperature of the Cosmic Microwave Background,
\href{https://doi.org/10.1088/0004-637X/707/2/916}{Astrophys. J. \textbf{707}, 916 (2009)}.
\href{https://arxiv.org/abs/0911.1955}{0911.1955}

\bibitem{WMAP:2010qai}
E.~Komatsu \textit{et al.} (WMAP),
Seven-Year Wilkinson Microwave Anisotropy Probe (WMAP) Observations: Cosmological Interpretation,
\href{https://doi.org/10.1088/0067-0049/192/2/18}{Astrophys. J. Suppl. \textbf{192}, 18 (2011)}.
\href{https://arxiv.org/abs/1001.4538}{1001.4538}

\bibitem{Torres:1997sn}
D.F.~Torres, H.~Vucetich and A.~Plastino, Early universe test of nonextensive statistics, 
\href{https://doi.org/10.1103/PhysRevLett.79.1588}{Phys. Rev. Lett. \textbf{79}, 1588 (1997)}. [erratum: \href{https://doi.org/10.1103/PhysRevLett.80.3889}{Phys. Rev. Lett. \textbf{80}, 3889 (1998)}].
\href{https://arxiv.org/abs/astro-ph/9705068}{astro-ph/9705068}

\bibitem{Lambiase:2011zz}
G.~Lambiase, Dark matter relic abundance and big bang nucleosynthesis in Horava's gravity, 
\href{https://doi.org/10.1103/PhysRevD.83.107501}{Phys. Rev. D \textbf{83}, 107501 (2011)}.
 
\bibitem{Lambiase:2012fv}
G.~Lambiase, Constraints on massive gravity theory from big bang nucleosynthesis, 
\href{https://doi.org/10.1088/1475-7516/2012/10/028}{J. Cosmol. Astropart. P. \textbf{10}, 028 (2012)}.
\href{https://arxiv.org/abs/1208.5512}{1208.5512}

\bibitem{Kneller} 
J.P.~Kneller and G.~Steigman, BBN for pedestrians, 
\href{https://doi.org/10.1088/1367-2630/6/1/117}{New J. Phys. \textbf{6}, 117 (2004)}.
\href{https://arxiv.org/abs/astro-ph/0406320}{astro-ph/0406320}

\bibitem{steigman07} 
G.~Steigman, Primordial nucleosynthesis in the precision cosmology era, 
\href{https://doi.org/10.1146/annurev.nucl.56.080805.140437}{Ann. Rev. Nucl. Part. Sci. \textbf{57}, 463 (2007)}.
\href{https://arxiv.org/abs/0712.1100}{0712.1100}

\bibitem{Steigman:2012ve} 
G.~Steigman, Neutrinos and big bang nucleosynthesis, 
\href{https://doi.org/10.1155/2012/268321}{Adv. High Energy Phys. \textbf{2012}, 268321 (2012)}. 
\href{https://arxiv.org/abs/1208.0032}{1208.0032}

\bibitem{Workman:2022ynf}
R.L.~Workman \textit{et al.} (Particle Data Group), Review of Particle Physics, 
\href{https://doi.org/10.1093/ptep/ptac097}{Prog. Theor. Exp. Phys. \textbf{2022}, 083C01 (2022)}.

 \bibitem{Aver:2015iza}
E.~Aver, K.A.~Olive and E.D.~Skillman, The effects of He~I $\lambda10830$ on helium abundance determinations, 
\href{https://doi.org/10.1088/1475-7516/2015/07/011}{J. Cosmol. Astropart. Phys. \textbf{07}, 011 (2015)}. 
\href{https://arxiv.org/abs/1503.08146}{1503.08146}
  
\bibitem{Peimbert} 
A.~Peimbert, M.~Peimbert and V.~Luridiana, The primordial helium abundance and the number of neutrino families, 
Rev. Mex. Astron. Astrofis. \textbf{52}, 419 (2016).
\href{https://arxiv.org/abs/1608.02062}{1608.02062}
 
\bibitem{Aver:2020fon}
E.~Aver, D.A.~Berg, K.A.~Olive, R.W.~Pogge, J.J.~Salzer and E.D.~Skillman, Improving helium abundance determinations with Leo P as a case study, 
\href{https://doi.org/10.1088/1475-7516/2021/03/027}{J. Cosmol. Astropart. Phys. \textbf{03}, 027 (2021)}. 
\href{https://arxiv.org/abs/2010.04180}{2010.04180}

\bibitem{Fields:2019pfx}
B.D.~Fields, K.A.~Olive, T.H.~Yeh and C.~Young, Big-Bang Nucleosynthesis after Planck, 
\href{https://doi.org/10.1088/1475-7516/2020/03/010}{J. Cosmol. Astropart. Phys. \textbf{03}, 010 (2020)}, [erratum: \href{https://doi.org/10.1088/1475-7516/2020/11/E02}{J. Cosmol. Astropart. Phys. \textbf{11}, E02 (2020)}].
\href{https://arxiv.org/abs/1912.01132}{1912.01132}

\bibitem{Bratt:2002xt}
J.D.~Bratt, A.C.~Gault, R.J.~Scherrer and T.P.~Walker,
Big Bang Nucleosynthesis constraints on brane cosmologies,
\href{https://doi.org/10.1016/S0370-2693(02)02637-0}{Phys. Lett. B \textbf{546}, 19-22 (2002)}.
\href{https://arxiv.org/abs/astro-ph/0208133}{astro-ph/0208133}

\bibitem{Abad} 
M.A.B.~Abad, G.M.~Tavares and M.~Schmaltz, Non-Abelian dark matter and dark radiation, \href{https://doi.org/10.1103/PhysRevD.92.023531}{Phys. Rev. D \textbf{92}, 023531 (2015)}. 
\href{https://arxiv.org/abs/1505.03542}{1505.03542}

\bibitem{Weinberg02} 
S.~Weinberg, Goldstone Bosons as Fractional Cosmic Neutrinos, 
\href{https://doi.org/10.1103/PhysRevLett.110.241301}{Phys. Rev. Lett. \textbf{110}, 241301 (2013)}. 
\href{https://arxiv.org/abs/1305.1971}{1305.1971}

\bibitem{Abazajian} 
K.N.~Abazajian {\it et al.}, Light Sterile Neutrinos: A White Paper,
\href{https://arxiv.org/abs/1204.5379}{1204.5379}


\bibitem{CMB-S4:2016ple}
K.N.~Abazajian \textit{et al.} (CMB-S4),
CMB-S4 Science Book, First Edition,
\href{https://arxiv.org/abs/1610.02743}{1610.02743}


\bibitem{CMB-S4:2022ght}
K.~Abazajian \textit{et al.} (CMB-S4),
Snowmass 2021 CMB-S4 White Paper,
\href{https://arxiv.org/abs/2203.08024}{2203.08024}

\bibitem{prodanovic} 
T.~Prodanović, G.~Steigman and B.D.~Fields, The deuterium abundance in the local interstellar medium, 
\href{https://doi.org/10.1111/j.1365-2966.2010.16734.x}{Mon. Not. R. Astron. Soc. \textbf{406}, 1108 (2010)}.
\href{https://arxiv.org/abs/0910.4961}{0910.4961}

\bibitem{Cooke:2017cwo}
R.J.~Cooke, M.~Pettini and C.C.~Steidel,
One Percent Determination of the Primordial Deuterium Abundance,
\href{https://doi.org/10.3847/1538-4357/aaab53}{Astrophys. J. \textbf{855}, 102 (2018)}.
\href{https://arxiv.org/abs/1710.11129}{1710.11129}

\bibitem{Mossa:2020gjc}
V.~Mossa, K.~St\"ockel, F.~Cavanna, F.~Ferraro \textit{et al.},
The baryon density of the Universe from an improved rate of deuterium burning,
\href{https://doi.org/10.1038/s41586-020-2878-4}{Nature \textbf{587}, 210 (2020)}.

\bibitem{Pitrou:2020etk}
C.~Pitrou, A.~Coc, J.P.~Uzan and E.~Vangioni,
A new tension in the cosmological model from primordial deuterium?,
\href{https://doi.org/10.1093/mnras/stab135}{Mon. Not. Roy. Astron. Soc. \textbf{502}, 2474 (2021)}.
\href{https://arxiv.org/abs/2011.11320}{2011.11320}

\bibitem{Pisanti:2020efz}
O.~Pisanti, G.~Mangano, G.~Miele and P.~Mazzella,
Primordial Deuterium after LUNA: concordances and error budget,
\href{https://doi.org/10.1088/1475-7516/2021/04/020}{J. Cosmol. Astropart. Phys. \textbf{04}, 020 (2021)}.
\href{https://arxiv.org/abs/2011.11537}{2011.11537} 

\bibitem{Yeh:2020mgl}
T.H.~Yeh, K.A.~Olive and B.D.~Fields,
The impact of new $d(p,\gamma)$3 rates on Big Bang Nucleosynthesis,
\href{https://doi.org/10.1088/1475-7516/2021/03/046}{J. Cosmol. Astropart. Phys. \textbf{03}, 046 (2021)}.
\href{https://arxiv.org/abs/2011.13874}{2011.13874} 



\end{thebibliography}
\end{document}